\documentclass[%
11pt,
reprint,
onecolumn,
tightenlines,
superscriptaddress,
preprintnumbers,
nofootinbib,
amsmath,amssymb,amsthm,
physrev,
eqsecnum,tikz,
]{revtex4-2}

\usepackage{isomath}
\usepackage{amsmath,amsthm}
\usepackage{amsbsy}
\usepackage{amssymb}
\usepackage{amscd}
\usepackage{amsfonts}
\usepackage{stmaryrd}
\usepackage{siunitx}
\usepackage{euscript}
\usepackage[utf8]{inputenc}
\usepackage[T1]{fontenc}
\usepackage{newtxtext} 
\everymath{\displaystyle}
\usepackage{exscale}

\usepackage{graphicx}
\usepackage{boxedminipage}
\usepackage{calc}
\usepackage[usenames,dvipsnames]{xcolor}
\graphicspath{ {media/} }
\usepackage[caption=false,justification=centerlast]{subfig}

\usepackage{setspace}
\usepackage{enumitem}
\setitemize{noitemsep,topsep=0pt,parsep=0pt,partopsep=0pt}
\setenumerate{noitemsep,topsep=0pt,parsep=0pt,partopsep=0pt}
\setdescription{noitemsep,topsep=0pt,parsep=0pt,partopsep=0pt}

\usepackage{hyperref}

\usepackage[small]{titlesec}

\titlespacing*{\section}{0pt}{12pt plus 4pt minus 2pt}{2pt plus 2pt minus 2pt}
\titlespacing*{\subsection}{0pt}{12pt plus 4pt minus 2pt}{2pt plus 2pt minus 2pt}
\titlespacing*\subsubsection{0pt}{12pt plus 4pt minus 2pt}{2pt plus 2pt minus 2pt}
\titlespacing*\paragraph{0pt}{12pt plus 4pt minus 2pt}{2pt plus 2pt minus 2pt}

\makeatletter

    \renewcommand*{\p@subsection}{}
    
    \renewcommand*{\p@subsubsection}{}
\makeatother

\usepackage{isomath}
\usepackage{amsmath}
\usepackage{amssymb}
\usepackage{amscd}
\usepackage{amsfonts}

\newcommand{\beq}{\begin{equation}}
\newcommand{\eeq}{\end{equation}}
\newcommand{\beqs}{\begin{eqnarray}}
\newcommand{\eeqs}{\end{eqnarray}}
\newcommand{\half}{\frac{1}{2}}

\newcommand{\trace}{\mathop{\rm tr}\nolimits}

\newcommand{\parderiv}[2]{\frac{\partial #1}{\partial #2}}
\newcommand{\dm}{\ \mathrm{d}}

\newcommand{\bfe}{{\mathbold e}}

\newcommand{\bfn}{{\mathbold n}}

\newcommand{\bfq}{{\mathbold q}}
\newcommand{\bfr}{{\mathbold r}}

\newcommand{\bft}{{\mathbold t}}
\newcommand{\bfu}{{\mathbold u}}
\newcommand{\bfv}{{\mathbold v}}

\newcommand{\bfx}{{\mathbold x}}

\newcommand{\bfF}{{\mathbold F}}

\newcommand{\bfI}{{\mathbold I}}

\newcommand{\bfR}{{\mathbold R}}

\newcommand{\bfX}{{\mathbold X}}

\newcommand{\mrn}{{\mathring{n}}}
\newcommand{\mrbfn}{{\mathbold {\mathring{n}}}}
\newcommand{\mrbe}{{\bar{\beta}_e}}

\newcommand{\mrtheta}{{\mathring{\theta}}}
\newcommand{\thetaTop}{\theta_{\mathsf{top}}}

\newcommand{\Wvol}{\mathcal{W}_{\mathsf{vol}}}
\newcommand{\WnH}{\mathcal{W}_{\mathsf{nH}}}
\newcommand{\Eddd}{\mathcal{E}_{\mathsf{3D}}}
\newcommand{\Omddd}{\Omega_{\mathsf{3D}}}
\newcommand{\Edd}{\mathcal{E}_{\mathsf{2D}}}
\newcommand{\Omdd}{\Omega_{\mathsf{2D}}}

\newcommand{\eOneLoc}{\bfe_1^{\mathsf{loc}}}
\newcommand{\eTwoLoc}{\bfe_2^{\mathsf{loc}}}
\newcommand{\eThreeLoc}{\bfe_3^{\mathsf{loc}}}

\newcommand{\yloc}{y^{\mathsf{loc}}}
\newcommand{\zloc}{z^{\mathsf{loc}}}
\newcommand{\bfxloc}{\bfx^{\mathsf{loc}}}
\newcommand{\bfxloch}{\bfx^{\mathsf{loc},h}}
\newcommand{\M}{\mathsf{M}}

\begin{document}

%%%%%%%%%%%%%%%%%%%%%
%%%%%%%%%%%%%%%%%%%%%
%%%%%%%%%%%%%%%%%%%%%
%%%%%%%%%%%%%%%%%%%%%

\preprint{To appear in Soft Matter (\url{doi.org/10.1039/D3SM00664F})}

\title{A Dimensionally-Reduced Nonlinear Elasticity Model for Liquid Crystal Elastomer Strips with Transverse Curvature}

\author{Kevin LoGrande}
    \email{klogrand@andrew.cmu.edu}
    \affiliation{Department of Civil and Environmental Engineering, Carnegie Mellon University}
    \affiliation{Computation and Information Sciences Directorate, CCDC Army Research Lab}

\author{M. Ravi Shankar}
    \affiliation{Department of Industrial Engineering, University of Pittsburgh}

\author{Kaushik Dayal}
    \affiliation{Department of Civil and Environmental Engineering, Carnegie Mellon University}
    \affiliation{Center for Nonlinear Analysis, Department of Mathematical Sciences, Carnegie Mellon University}
    \affiliation{Department of Mechanical Engineering, Carnegie Mellon University}

\date{\today}

%%%%%%%%%%%%%%%%%%%%%
%%%%%%%%%%%%%%%%%%%%%
%%%%%%%%%%%%%%%%%%%%%
%%%%%%%%%%%%%%%%%%%%%

%%%%%%%%%%%%%%%%%%%%%
%%%%%%%%%%%%%%%%%%%%%
%%%%%%%%%%%%%%%%%%%%%
%%%%%%%%%%%%%%%%%%%%%

\begin{abstract}
	
    Liquid Crystalline Elastomers (LCEs) are active materials that are of interest due to their programmable response to various external stimuli such as light and heat. 
    When exposed to these stimuli, the anisotropy in the response of the material is governed by the nematic director, which is a continuum parameter that is defined as the average local orientation of the mesogens in the liquid crystal phase.
    This nematic director can be programmed to be heterogeneous in space, creating a vast design space that is useful for applications ranging from artificial ligaments to deployable structures to self-assembling mechanisms. 
    Even when specialized to long and thin strips of LCEs -- the focus of this work -- the vast design space has required the use of numerical simulations to aid in experimental discovery. 
    To mitigate the computational expense of full 3-d numerical simulations, several dimensionally-reduced rod and ribbon models have been developed for LCE strips, but these have not accounted for the possibility of initial transverse curvature, like carpenter's tape spring.
    Motivated by recent experiments showing that transversely-curved LCE strips display a rich variety of configurations, this work derives a dimensionally-reduced 1-d model for pre-curved LCE strips.
    The 1-d model is validated against full 3-d finite element calculations, and it is also shown to capture experimental observations, including tape-spring-like localizations, in activated LCE strips.
\end{abstract}

\maketitle

%%%%%%%%%%%%%%%%%%%%%
%%%%%%%%%%%%%%%%%%%%%
%%%%%%%%%%%%%%%%%%%%%
%%%%%%%%%%%%%%%%%%%%%
\section{Introduction}

Liquid crystalline elastomers (LCEs) are active materials that have grown in popularity due to their proposed use in a wide variety of applications, from artificial muscles, to deployable structures, to soft robotics \cite{warner1996nematic,white2015programmable,fuchi2015topology,babaei2021torque,ahn2016photoinduced,ware2016localized,ahn2019bioinspired,ware2015voxelated,babaei2017steering,dradrach2023light}.
LCEs, at the molecular level, are made of ellipsoidal molecules (mesogens) typical to the liquid crystal phase, which are further cross-linked by polymer chains. 
The average direction of alignment of the mesogens at a material point is called the nematic director. 
When exposed to external stimuli---depending on the molecular makeup of the mesogen, one could use heat, light, or electric field---LCEs will shrink in the direction of the nematic director and expand in transverse directions. Because the nematic director is highly programmable in the manufacturing of LCEs, a wide range of controlled deformations is possible \cite{ware2015voxelated,javed2022programmable,lewis2023programming}.

Even restricting the design space to the mechanical deformation of long and thin LCE strips, which are the focus of the present work, allows for a large range of deformation mechanisms. 
Varying the nematic director through the thickness of these strips (e.g., in a twisted or splay-bend geometry) can lead to the spontaneous development of transverse curvature, buckling, instabilities, and coupled bending/twisting behavior, e.g. \cite{CLEMENT2021101362,giudici2022curvature}. 
However, this highly programmable nature of LCEs necessitates the use of modeling to explore the vast space of design parameters. 
Fully 3-d models based on nonlinear elasticity can be applied to strips by treating them as slender 3-d continuum bodies, and have been shown to perform well when numerical methods are carefully designed, e.g. \cite{kadapa2021advantages}, but these can be computationally expensive.
The present work aims to leverage the small width and thickness dimensions to derive a dimensionally-reduced 1-d ribbon model that can capture the complex 3-d deformations of LCE strips.
In particular, it accounts for the transverse curvature that provides a rich space of configurations \cite{CLEMENT2021101362}.

\paragraph*{Prior Work.} 

Several prior contributions have put forward energy densities for dimensionally-reduced LCE structures such as strips, e.g. \cite{agostiniani2017dimension,Singh2022,D0SM00642D,bartels2022modeling,agostiniani2011gamma,agostiniani2017shape,agostiniani2020rigorous}, but these have not considered the presence of transverse curvature.
Recent experiments, particularly \cite{CLEMENT2021101362}, however, observe interesting phenomena in LCE strips with transverse curvature: e.g., high torque-density snap-through in pre-curved strips, which is relevant to the use of LCEs as actuators \cite{babaei2021torque}, and motivate our work.

Further, both flat and pre-curved strips and rods have long been studied, and the slender geometry leads to large nonlinear deformations and a rich variety of instabilities, e.g. \cite{zajac1962stability,green1936equilibrium,green1937elastic,purohit2008plectoneme,steigmann2008two,korte2011triangular,fosdick2016mechanics,yu2019bifurcations,dondl2023efficient,radisson2023dynamic,radisson2023elastic}.
Particularly in the context of pre-curved strips, a ubiquitously-observed instability is the classic tape spring instability \cite{antman2005problems,audoly2000elasticity,seffen2000folding}.
While there have been several contributions that put forward dimensionally-reduced models of strips and related structures that account for the tape-spring and other instabilities, these are in the context of mechanical materials without the consideration of active behavior, e.g. \cite{kumar2023asymptotic,audoly2016buckling}.
As observed in \cite{CLEMENT2021101362}, there is a rich interplay between the geometric feature of transverse curvature in the tape spring geometry and the stimuli-driven spontaneous deformations that are possible in LCE. 
There has also been recent work devoted to dimension reduction for models including differential growth for rods with a general cross-sectional shape\cite{moulton2020morphoelastic}, but these models define the cross-section by a single small parameter, i.e., they consider \textit{rods}, not \textit{strips}. 
The observations of interest from \cite{CLEMENT2021101362} require thin strips with a hierarchy of dimensions, i.e., the thickness is much smaller than the width, which is in turn much smaller than the length. 
Accounting for this hierarchy of scales enables our model to capture highly localized strains like tape-spring instabilities.

\paragraph*{Contributions of This Paper.}

Several important elements of our approach follow \cite{GUINOT201273}, with some notable differences: first, that work begins from 2-d and reduces to 1-d, whereas we begin from 3-d and reduce to 1-d; second, that work considers a purely-mechanical response, whereas we consider the effect of stimuli-driven deformation; and third, 
that work starts with an isotropic small-strain plate model, whereas we use a nonlinear elastic model in 3-d.

We highlight here that we cannot simply adapt the model and approach of \cite{GUINOT201273} to our setting.
Rather, we must necessarily start from 3-d because we consider twisted nematics -- i.e., the nematic director varies through the thickness -- which is the root cause of the observed complexity.
Further, the physics of LCE materials requires a nonlinear elastic model to capture the range of observed behavior.

We start by assuming an ansatz on the full 3-d deformation field that is inspired by the 2-d membrane ansatz proposed in \cite{GUINOT201273}. 
Further, our ansatz also accounts for the variation of the nematic director through the thickness of the strip.
Furthermore, the warping function accounts for the transverse shear that we expect during nematic activation of an LCE strip with a director field that varies through the thickness. 
Our proposed kinematic ansatz is used in combination with the standard neo-Hookean strain energy density adapted to glassy LCE \cite{anderson1999continuum,desimone2009elastic}.
Following the conventional procedure of integrating the energy density over the cross-section of the strip, we arrive at a reduced lineic strain energy density that governs the deformation of the centerline of the strip while tracking changes in the shape of the cross-section. 

We test the reduced model by comparing its predictions against those of the full 3-d model in examples of interest to experiment.
We then use the model to predict and explain recent experimental observations from \cite{CLEMENT2021101362}. 
The success of our dimensionally-reduced model is also notable in the computational efficiency. 
We highlight a powerful framework for a very general class of cross-sectional deformations proposed in \cite{le2023numerical}.
However, that approach also requires intensive computational effort: it requires either a 2-d finite element (FE) calculation for every element along the centerline curve (i.e., an FE\textsuperscript{2} method), or it requires the training of the model on an arbitrarily large class of cross-section deformations to try to learn an effective behavior for implementation into a 1-d model.
While potentially much more accurate, both of these approaches are far more computationally intensive than the simple 1-d FE calculations required for the model proposed in this paper. 
This fast computation is a vital component of our work, since it enables a rapid exploration of the design space needed to guide experimental design.

\paragraph*{Organization.}
In Section \ref{sec:constitutive}, we provide the 3-d formulation, nematic patterning, and constitutive model that completely describe the 3-d pre-curved LCE strip.
In Section \ref{sec:kinematics}, we develop the ansatz for the deformation. 
In Section \ref{sec:reduction}, we apply the ansatz to obtain the dimension reduction of the full 3-d model to the reduced 1-d model.
Finally, in Sections \ref{sec:num-mthd} and \ref{sec:results}, we describe, respectively, the numerical solution procedure and the results from the dimensionally-reduced model.

%%%%%%%%%%%%%%%%%%%
%%%%%%%%%%%%%%%%%%%
%%%%%%%%%%%%%%%%%%%
%%%%%%%%%%%%%%%%%%%
\section{Three-dimensional Formulation, Nematic Patterning, and Constitutive Modeling}
\label{sec:constitutive}

\subsection{Elastic Strain Energy Density}

 It is usual to model LCEs using the neo-Hookean hyperelasticity model, which is based on the statistical mechanics of polymer chains, but with the stress-free state dependent on the state of the nematic director.
 Our model described here is standard in the literature \cite{desimone2009elastic}.

Given a deformation gradient tensor, $\bfF$, the compressible neo-Hookean strain energy density is:
\begin{equation} \label{eq:Neo_Hookean}
    \WnH\left(\bfF\right)=\frac{\mu}{2}\left(\trace\left(\bfF^T\bfF\right)-3-2\log J\right)+\frac{\kappa}{2}\Wvol\left(J\right)\quad \text{where } J=\det\bfF
\end{equation}
where $\mu$ and $\kappa$ are positive material constants with the dimension of energy per unit volume, and correspond to the shear and bulk moduli respectively. 
$\Wvol\left(J\right)$ term accounts for compressibility effects, and satisfies that: (1) $\Wvol\left(J\right)=0\iff J=1$; (2) As $J\to0^+$, $\Wvol\left(J\right)\to\infty$; and (3) $\Wvol^{''}\left(1\right)>0$.
For the present work, we will use:
\begin{equation} \label{eq:wvol}
    \Wvol\left(J\right)=J^2-1-2\log\left(J\right)
\end{equation}
The effect of the thermal or optical stimulus is to change the nematic ordering, that we denote \textit{nematic activation}.
In the thermal case, at low temperature the nematic mesogens have local ordering that can be described by the unit vector director field $\bfn$; at higher temperatures, the nematic ordering is lost and becomes isotropic.
In the optical case, the LCE specimens are doped with azobenzene -- a light-sensitive molecule -- that undergoes a \textit{trans-cis} transformation when exposed to light of the appropriate wavelength \cite{BERG20145849}. 
While the molecular-scale mechanisms are different in the thermal and optical cases, at the continuum scale both can be described by a stress-free deformation that we will denote $\bfF_l$.
A key difference between the thermal and optical cases is that the former can be well-approximated as having a uniform temperature -- and consequently uniform nematic activation -- in the body, whereas absorption of light through the thickness is essential to model the optical case accurately, e.g. \cite{babaei2017steering,babaei2021torque}.
In this work, we assume uniform nematic activation.

We can write $\bfF_l$ using the spectral representation as:
\begin{equation} \label{eq:F_l_defn}
    \bfF_{l}=\alpha^{1/3}\bfn\otimes\bfn+\alpha^{-1/6}\left(\bfI-\bfn\otimes\bfn\right)
\end{equation}
where $\bfn$ is the nematic director of unit length, $\bfI$ is the identity tensor, and $\alpha>0$ is the coefficient of stretch along the nematic direction.
Notice that $\det\bfF_l = 1$.

In the present work, we take our reference to be the low-temperature state.
The nematic activation at high temperature results in contraction along the director and isotropic expansion in the plane orthogonal to the director. 
Consequently, we have $\alpha < 1$.
We note that our convention is opposite to others assumed in the literature, e.g. \cite{agostiniani2017dimension} assumes a high-temperature reference which merely implies that $\alpha > 1$ for them.

Given $\bfF_l$ from \eqref{eq:F_l_defn}, we can define the usual multiplicative decomposition of $\bfF$ in terms of the elastic deformation $\bfF_e$:
\begin{equation} \label{eq:F_e_defn}
    \bfF=\bfF_e \bfF_{l} \quad \Longleftrightarrow \quad \bfF_e=\bfF \bfF_{l}^{-1}
\end{equation}
Noting the spectral decomposition of $\bfF_l$, we can write $\bfF_e$ as:
\begin{equation} \label{eq:F_e_calc}
    \bfF_e=\alpha^{1/6}\bfF+\left(\alpha^{-1/3}-\alpha^{1/6}\right)\bfF\left(\bfn\otimes\bfn\right)
\end{equation}
Since $\bfF_l$ changes the stress-free state, the elastic energy density for the LCE can be written:
\begin{equation} \label{eq:Neo_F_e}
    \hat{\mathcal{W}}\left(\bfF\right)
    = \WnH\left(\bfF_e\right) 
\end{equation}

%%%%%%%%%%%%%%%%%%%
%%%%%%%%%%%%%%%%%%%
%%%%%%%%%%%%%%%%%%%
%%%%%%%%%%%%%%%%%%%
\subsection{Change of Reference to a Flat Strip}
\label{sec:change-ref}

\begin{figure}[htb!]
    \centering
    \includegraphics[width=10cm]{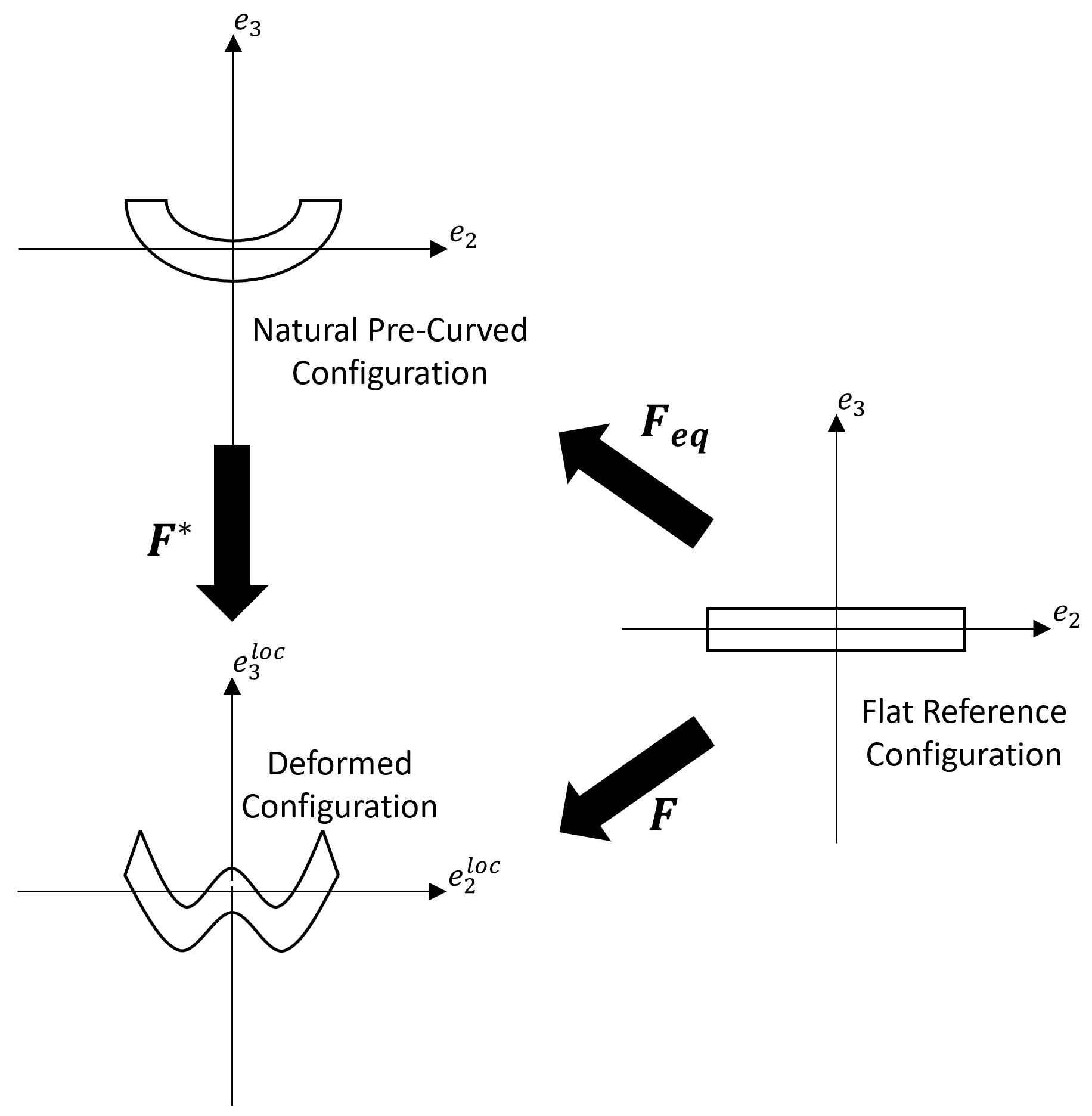}
    \caption{Change of reference to an initially flat configuration.\label{fig:refChange}}
\end{figure}

The strips that we consider in this work are produced as nominally flat, but then annealed to significantly reduce the stresses.
In the process of annealing, the strips also develop small, but critical, transverse curvature; this is achieved by adhering the strip lengthwise to a thin cylinder during annealing  \cite{CLEMENT2021101362}.
Given this process of annealing, we assume in this work that the annealed configuration with transverse curvature is stress-free, i.e., the natural state.

Working directly with this pre-curved natural reference poses some difficulties, in terms of algebraic complexity, for deriving a dimensionally-reduced model.
Therefore, we use a non-stress-free flat strip as our reference configuration. 
We discuss here the transformation of the reference configuration, as shown in Figure \ref{fig:refChange}.

Following Figure \ref{fig:refChange}, we have the relation:
\begin{equation} \label{eq:refChange}
    \bfF^*=\bfF\bfF_{eq}^{-1}
\end{equation}
Note that if the strip is not pre-curved, $\bfF_{eq}=\bfI$ and we recover $\bfF^*=\bfF$.

Since the deformation from the natural stress-free configuration to the current configuration is $\bfF^*$, it follows that we should replace \eqref{eq:F_e_defn} by:
\begin{equation} \label{eq:F*_e_defn}
    \bfF^*=\bfF_e\bfF_{l}
\end{equation}

Combining \eqref{eq:refChange} and \eqref{eq:F*_e_defn} gives the expression for the elastic deformation using the flat reference configuration:
\begin{equation}\label{eq:F_e_defn_flat}
    \bfF_e=\bfF\bfF_{eq}^{-1}\bfF_l^{-1}    
\end{equation}
which should be used as the argument for the neo-Hookean energy density in \eqref{eq:Neo_F_e}.

%%%%%%%%%%%%%%%%%%%
%%%%%%%%%%%%%%%%%%%
%%%%%%%%%%%%%%%%%%%
%%%%%%%%%%%%%%%%%%%
\subsection{Director Patterning through the Cross-Section}
\label{sec:director}

\begin{figure}[htb!]
    \centering
    \includegraphics[width=10cm]{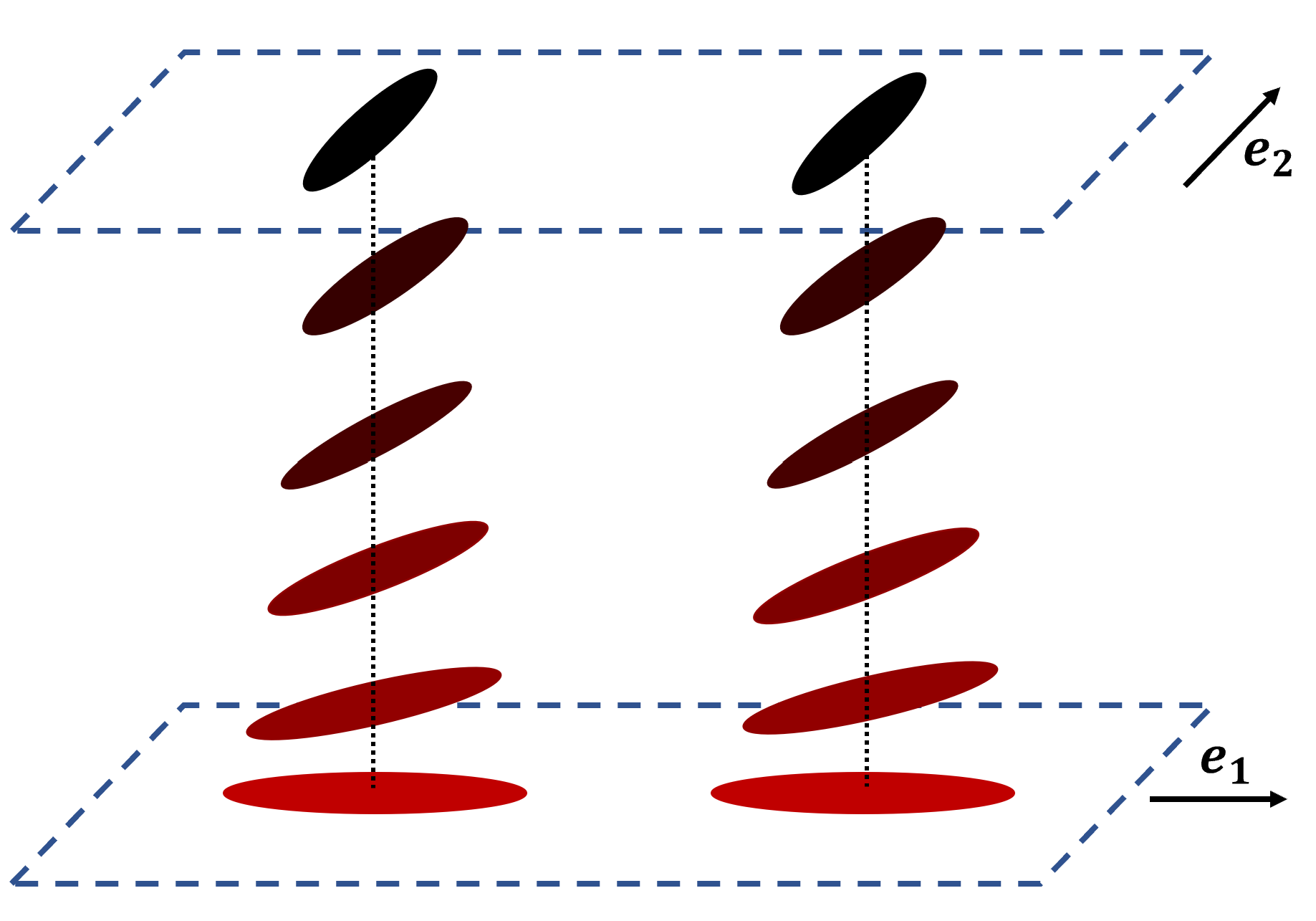}
    \caption{A twisted nematic director pattern in the flat reference state. The length of the strip is aligned along $\bfe_1$, the width along $\bfe_2$, and the thickness along $\bfe_3$. Note that $\theta_{top}=\pi/2$ here, corresponding to $\mrbfn=\bfe_2$ on the top surface and $\mrbfn=\bfe_1$ on the bottom surface.
    \label{fig:twist-nem}}
\end{figure}

We use in the current work a twisted nematic director pattern.
In the flat reference configuration, using the orthonormal frame shown in Figure \ref{fig:twist-nem}, this pattern has a director that lies in the $\bfe_1 - \bfe_2$ plane and which varies only through the thickness.
This director can be represented in the fixed orthonormal frame as:
\begin{equation} \label{eq:director_defn}
    \mrbfn=\cos{\mrtheta}\hspace{2pt}\bfe_1+\sin{\mrtheta}\hspace{2pt}\bfe_2
\end{equation}
where $\mrtheta$ varies linearly through the thickness of the strip.

The glassy behavior of the LCE causes the director to be fixed with respect to the polymer matrix background.
Therefore, it is assumed to be mapped under the deformation gradient as a material vector, but normalized to remain unit \cite{desimone2009elastic}:
\begin{equation} \label{eq:director_map}
    \bfn=\frac{\bfF\mrbfn}{|\bfF\mrbfn|}
\end{equation}
Note that since $\mrbfn$ is defined in the flat reference, it is mapped under $\bfF$ and not $\bfF^*$. 

We note that our approach can also be used easily for other nematic director patterns. 
One popular configuration is the \textit{splay-bend} pattern
which lies in the $\bfe_1 - \bfe_3$ plane with the nematic director normal to the surface at the top of the strip and parallel to the centerline at the bottom of the strip, with linear variation through the thickness. 
We do, however, in the dimension reduction procedure in Section \ref{sec:reduction} assume that the only variation of the director is in the thickness direction.

%%%%%%%%%%%%%%%%%%%
%%%%%%%%%%%%%%%%%%%
%%%%%%%%%%%%%%%%%%%
%%%%%%%%%%%%%%%%%%%
\section{Kinematic Assumptions for Dimension Reduction}
\label{sec:kinematics}

To conduct the dimension reduction procedure, we follow the approach of constructing an ansatz for the deformation that will enable us to integrate over the width and thickness directions.
Our ansatz is inspired by \cite{GUINOT201273}, but extended to the 3-d setting.

Consider a long, thin, and initially flat strip whose body in the reference state occupies the space, $\Omddd$, represented by the coordinates $(s_1,s_2,s_3)\in[0;L]\times[-a/2,a/2]\times[-h/2,h/2]$ that are orthogonal and Cartesian in the reference configuration, but are mapped to curvilinear coordinates in the deformed configuration. 
Here, $L$ is the length of the centerline of the strip, $a$ is the length of the cross-section curve of the middle surface, and $h$ is the uniform thickness of the strip. 
We construct a fixed orthonormal frame $(O,\bfe_1,\bfe_2,\bfe_3)$ from the initially flat configuration, such that $O$ is the centroid of the cross-section at $s_1=0$; $(O,\bfe_1)$ contains the centerline of the strip; and $\bfe_2$ and $\bfe_3$ align with the directions along the width and thickness of the strip respectively. 
A material point of the strip in the flat reference configuration has the position $\bfX$ given by the expression:
\begin{equation} \label{eq:ref_posn}
    \bfX\left(s_1,s_2,s_3\right)=s_1\bfe_1+s_2\bfe_2+s_3\bfe_3.
\end{equation}

\begin{figure}[htb!]
    \subfloat[]{\includegraphics[width=0.47\textwidth]{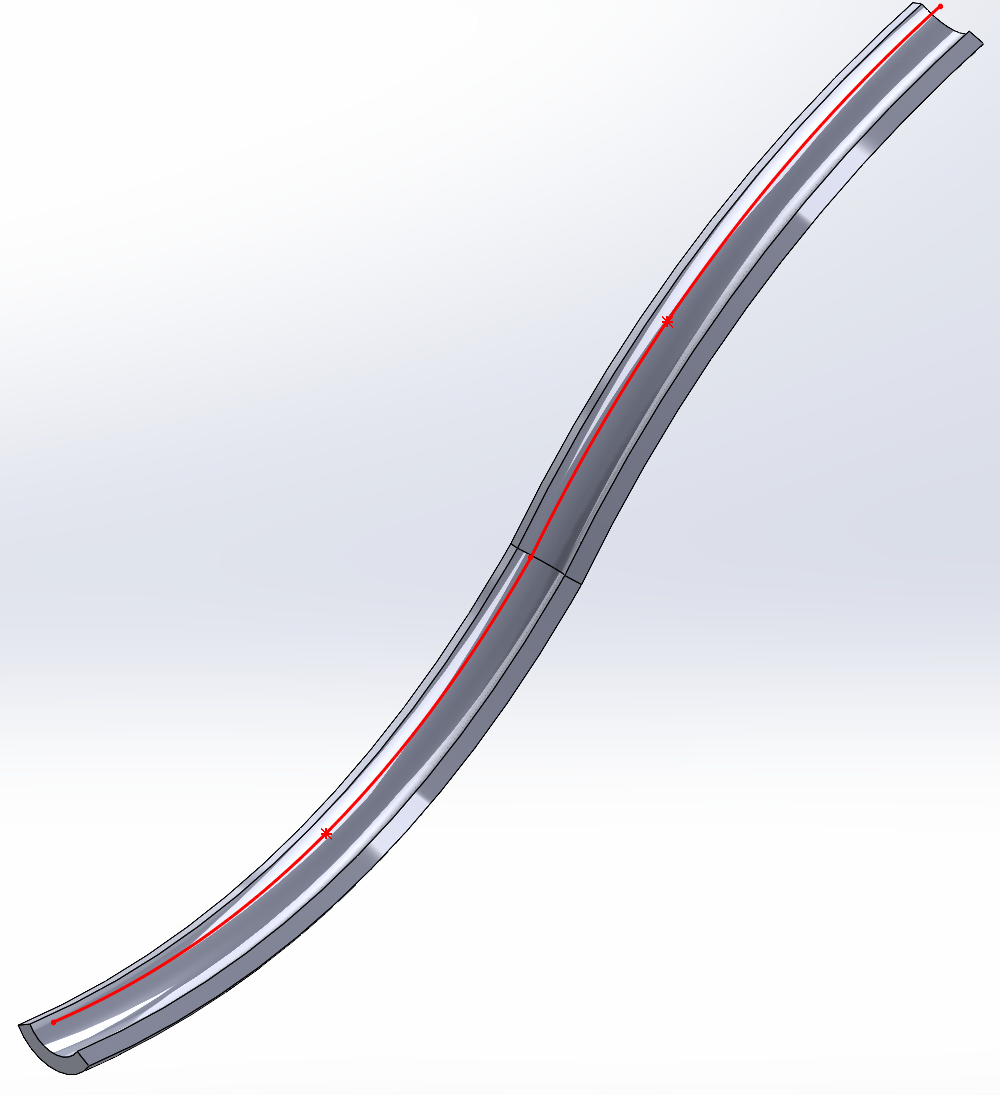}}
    \hfill
    \subfloat[]{\includegraphics[width=0.47\textwidth]{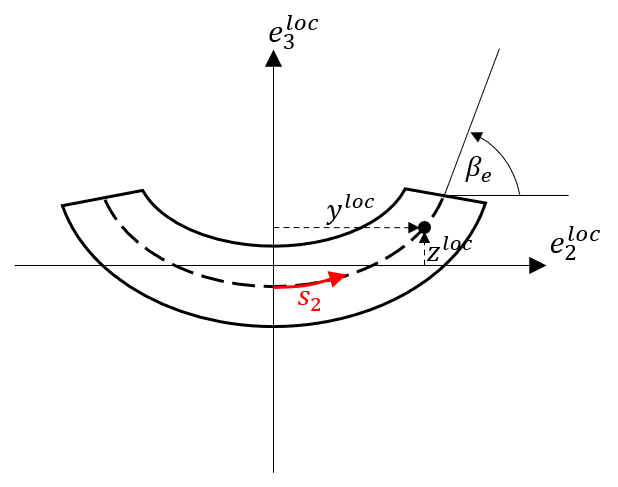}}
    \caption{Schematic of an LCE strip in the deformed configuration. The red line in the image of the full strip (left) corresponds to the $s_1$ coordinate direction. Each point along this centerline curve will have a corresponding position coordinate, $\bfu(s_1)$, and a rotation matrix, $\bfR(s_1)$ that track the local orthonormal frame at each point. A typical deformed cross-section is shown (right) with the open circular middle plane shown with the dashed line and the $s_2$ coordinate direction in red. The coordinates $\yloc$ and $\zloc$ track the position of the $s_2$ coordinate, while the $s_3$ coordinate will traverse through the thickness of the strip in the direction perpendicular to the cross-section curve.} 
    \label{fig:x-sctn}
\end{figure}

In the deformed configuration, we decompose the position vector of each material point into a rigid motion measured in the global coordinate frame and a local deformation of the cross-section, measured in a local orthonormal frame, $(\eOneLoc,\eTwoLoc,\eThreeLoc)$, that is related to the fixed orthonormal frame $(\bfe_1, \bfe_2, \bfe_3)$ as follows:
\begin{equation} \label{eq:local_basis}
   \bfe_i^{\mathsf{loc}}(s_1)=\bfR(s_1)\bfe_i
\end{equation}
The rigid motion of each cross-section is given by a displacement, $\bfu(s_1) = u_1(s_1) \bfe_1 + u_2(s_1)\bfe_2 + u_3(s_1)\bfe_3$, and a finite rotation characterized by a rotation tensor, $\bfR(s_1)$. The local cross-section deformation is described by the coordinates, $\yloc(s_1,s_2)$ and $\zloc(s_1,s_2)$, which track the position of the $s_2$ coordinate on the middle surface cross-section curve in the local frame (Fig. \ref{fig:x-sctn}). We further introduce the $s_3$ coordinate that traverses through the thickness of the strip. Since this direction remains perpendicular to the middle surface cross-section curve, it can be expressed in the local frame by components represented by the derivatives of $\yloc$ and $\zloc$ with respect to $s_2$, normalized by a local stretch factor, $j_2(s_1,s_2)$ given by the following:
\begin{equation} \label{eq:j2}
    j_2(s_1,s_2)=\sqrt{\left(\yloc_{,2}\right)^2+\left(\zloc_{,2}\right)^2}.
\end{equation}
where $(\bullet)_{,i}$ refers to the the partial derivative of $(\bullet)$ with respect to $s_i$. Lastly, we define the out-of-plane warping displacement of the cross-section along the length of the strip, $\chi(s_1,s_2)$, described in detail below in Section \ref{sec:warping}.

Putting all this together, we have that the material point $\bfX(s_1, s_2, s_3)$ in the flat reference goes to the location $\bfx(s_1,s_2,s_3)$ in the deformed configuration, given by:
\begin{equation} \label{eq:def_posn}
    \bfx(s_1,s_2,s_3)
    =
    \underbrace{(s_1+u_1)\bfe_1+u_2\bfe_2+u_3\bfe_3}_{=:\tilde{\bfu}}
    +
    \underbrace{\chi\eOneLoc+\left(\yloc-\frac{\zloc_{,2}}{j_2}s_3\right)\eTwoLoc+\left(\zloc+\frac{\yloc_{,2}}{j_2}s_3\right)\eThreeLoc}_{=: \bfR^T\bfxloc}
\end{equation}

We note that the introduction here of the $s_3$ coordinate is representative of our three-dimensional generalization of the kinematics developed in \cite{GUINOT201273}. 
Among other things, explicitly including the thickness coordinate is necessary to allow for a nematic director that varies through the thickness as described in Section \ref{sec:director}. 
The inclusion of this parameter is essential to our model, since the through-thickness variation of the nematic director is the root cause of the spontaneous bending and twisting curvatures in the strip, which in turn further drive the large class of deformations possible with LCE strips. 
The breadth of nonlinear behavior driven by this through-thickness variation requires careful averaging and re-scaling that we describe in Section \ref{sec:reduction}.
For ease of notation, we have defined $\tilde{\bfu}(s_1)$ and $\bfxloc(s_1,s_2,s_3)$ as the vectors containing the terms measured in the global frame and local frame respectively; the presence of $\bfR$ is to map to the global frame.

We can write the deformation gradient as:
\begin{equation} \label{eq:def_F}
    \bfx=\tilde{\bfu}+\bfR^T\bfxloc
    \Rightarrow
    \bfF=\nabla\tilde{\bfu}+\nabla\left(\bfR^T\bfxloc\right) 
\end{equation}
Note that because our curvilinear coordinates, $(s_1,s_2,s_3)$, correspond to the material coordinates in the reference configuration, the derivatives used for the construction of the deformation gradient are taken with respect to these curvilinear coordinates.

%%%%%%%%%%%%%%%%%%%
%%%%%%%%%%%%%%%%%%%
%%%%%%%%%%%%%%%%%%%
%%%%%%%%%%%%%%%%%%%
\subsection{Further Simplifying Assumptions}
\label{sec:assumptions}
We here present a list of assumptions that together simplify the kinematics and allow us to carry through with the dimensionality reduction.

%%%%%%%%%%%%%%%%%%%
%%%%%%%%%%%%%%%%%%%
%%%%%%%%%%%%%%%%%%%
%%%%%%%%%%%%%%%%%%%
\subsubsection{A Long, Thin Strip}
We assume that the thickness of the strip is much smaller than its width, which is in turn much smaller than the length of the strip. 
The extreme aspect ratio of the cross-section allows us to use a specific warping function for thin-walled open cross-sections, which we describe in Section \ref{sec:warping}, while the disparity between width and length of the strip will be important in our dimension reduction procedure given in Section \ref{sec:reduction}, where we will use truncated Taylor expansions in terms of the small dimensionless quantities $h/L$ and $a/L$.
\begin{equation} \label{eq:long_thin}
    \frac{h}{a}\ll 1 \quad \text{ and } \quad \frac{a}{L}\ll 1
\end{equation}

%%%%%%%%%%%%%%%%%%%
%%%%%%%%%%%%%%%%%%%
%%%%%%%%%%%%%%%%%%%
%%%%%%%%%%%%%%%%%%%
\subsubsection{Open Circular Cross-Section Curve}
We next assume that the curve that makes up the cross-section of the middle surface of the strip remains open circular throughout the deformation. 
This means that we can write explicitly a form for $\yloc(s_1,s_2)$ and $\zloc(s_1,s_2)$ in terms of a new variable, $\beta(s_1,s_2)$, which describes the angle the tangent line to the cross-section curve makes with the $\eTwoLoc$ direction:
\begin{equation} \label{eq:beta_defn}
    \yloc_{,2}=\cos{\beta}\quad,\quad \zloc_{,2}=\sin{\beta}
\end{equation}

The assumption that the cross-section is circular implies that $\beta(s_1,s_2)$ is linear in $s_2$, and hence has the expression:
\begin{equation} \label{eq:beta_e_defn}
    \beta=2\beta_e \frac{s_2}{a}
\end{equation}
where $\beta_e(s_1)$ is the opening angle of the cross-section (Fig. \ref{fig:x-sctn}(b)).
We note that a more general cross-section shape would correspond to a more general functional dependence of $\beta$ on $s_2$.

Assuming this form give the following expressions for $\yloc(s_1,s_2)$ and $\zloc(s_1,s_2)$.
\begin{equation} \label{eq:yloc_defn}
    \yloc=\frac{a}{2\beta_e} \sin \left(2\beta_e \frac{s_2}{a}\right)
\end{equation}
\begin{equation} \label{eq:zloc_defn}
    \zloc=\frac{a}{2\beta_e^2} \left(\sin\left(\beta_e\right)-\beta_e\cos\left(2 \beta_e \frac{s_2}{a}\right)\right)
\end{equation}

This leads to further simplifications. We note first that:
\begin{equation}
    \left(\yloc_{,2}\right)^2+\left(\zloc_{,2}\right)^2=1 \implies j_2=1
\end{equation}
This is a significant simplification in the expressions \eqref{eq:def_posn} and \eqref{eq:def_F}.

We can use a similar characterization to simplify $\bfF_{eq}$.
For a strip in the natural, stress-free configuration with an open-circular cross-section curve of opening angle $\mrbe$, we can write the inverse of $\bfF_{eq}$ as follows:
\begin{equation*}
    \bfF_{eq}^{-1}=
    \begin{bmatrix}
    1 & 0 & 0\\
    0 & \frac{a \cos{\left(2\mrbe s_2/a\right)}}{a-2\mrbe s_3} & \frac{a \sin{\left(2\mrbe s_2/a\right)}}{a-2\mrbe s_3}\\
    0 & -\sin{\left(2\mrbe s_2/a\right)} & \cos{\left(2\mrbe s_2/a\right)}
    \end{bmatrix}.
\end{equation*}
Note that if the opening angle $\mrbe=0$, then we recover $\bfF_{eq}=\bfI$.

Further, the determinant of $\bfF_{eq}$ depends on $s_3$ and $a$:
\begin{equation} \label{eq:det_feq_inverse}
    \det\bfF_{eq}^{-1}=\frac{a}{a-2\mrbe s_3},
\end{equation}
but, since $s_3\in[-h/2,h/2]$ and the thickness of the strip is much less than its width from \eqref{eq:long_thin}, we have $\det\bfF_{eq}^{-1} \approx 1$, implying that:
\begin{equation} \label{eq:det_F*_simplified}
    \det\bfF^*=\det\bfF_{eq}^{-1}\det\bfF\approx\det\bfF
\end{equation}
%%%%%%%%%%%%%%%%%%%
%%%%%%%%%%%%%%%%%%%
%%%%%%%%%%%%%%%%%%%
%%%%%%%%%%%%%%%%%%%
\subsubsection{Assumptions on the Warping}
\label{sec:warping}

Our final assumptions concern the form of the warping.
We assume that the out-of-plane displacement is due only to warping, and the warping is due only to twisting, following the classical Vlassov theory \cite{vlassov1962pieces} for the warping of thin-walled and open cross-sections.
This warping displacement can be written as:
\begin{equation} \label{eq:xw_defn}
    \chi=\omega k_t^r,
\end{equation}
where $\omega(s_1,s_2)$ is the so-called sectorial coordinate of the middle surface cross-section curve and $k_t^r(s_1)=(\eTwoLoc)_{,1}\cdot\eThreeLoc$ is the twisting curvature. Under the assumption that the thickness of the strip is much smaller than the width, $\omega(s_1,s_2)$ can be obtained from the following relation \cite{GUINOT201273}:
\begin{equation} \label{eq:omega_derive}
    \omega_{,2}=\yloc_{,2}\zloc-\yloc\zloc_{,2}\quad,\quad\omega=\int\omega_{,2}\dm s_2
\end{equation}
After integrating with respect to $s_2$, the constant of integration can be solved for by noting:
\begin{equation}
    \int_{-a/2}^{a/2}\omega \dm s_2=0.
\end{equation}
This gives us the following final form for $\omega(s_1,s_2)$:
\begin{equation} \label{eq:omega_defn}
    \omega=\frac{a}{4\beta_e^3}\left(a\sin{(\beta_e)}\sin{\left(2\beta_e\frac{s_2}{a}\right)}-2\beta_e^2s_2\right)
\end{equation}

Further, since the warping displacement, $\chi(s_1,s_2)$, remains small with respect to the width of the strip, we have that:
\begin{equation} \label{eq:inext_xSect}
    (\chi_{,2})^2+(\yloc_{,2})^2+(\zloc_{,2})^2\approx1  
\end{equation}
In other words, assuming that the cross-section is open circular constrains our middle surface cross-section curve to be effectively inextensible, which is reasonable for the deformations of long, thin strips. 

%%%%%%%%%%%%%%%%%%%
%%%%%%%%%%%%%%%%%%%
%%%%%%%%%%%%%%%%%%%
%%%%%%%%%%%%%%%%%%%
\section{Dimensionally-Reduced Model}
\label{sec:reduction}

Starting from \eqref{eq:Neo_F_e} and taking into account the approximate relation in \eqref{eq:det_F*_simplified}, we can rewrite the 3-d energy density as follows:
\begin{equation} \label{eq:enDen_index}
\begin{split}
    \hat{\mathcal{W}} &= \frac{\mu}{2}\left(\alpha^{1/3} F^*_{ij}F^*_{ij}+\left(\alpha^{-2/3}-\alpha^{1/3}\right)F^*_{ij}n_jF^*_{ik}n_k-3-2\log J\right)+\frac{\kappa}{2}\Wvol(J)\\
    &= \frac{\mu}{2}\left(\alpha^{1/3} F^*_{ij}F^*_{ij}+\left(\alpha^{-2/3}-\alpha^{1/3}\right)F_{ij}\mrn_jF_{ik}\mrn_k-3-2\log J\right)+\frac{\kappa}{2}\Wvol(J)
\end{split}
\end{equation}
where we use the Einstein summation convention here and in the following. 
Note that the second version of $\hat{\mathcal{W}}$ arises from the fact that $\mrbfn$ is defined in the flat reference state. 
Our 3-d energy, $\Eddd$, is the integral over the body, $\Omddd$, of the strain energy density:
\begin{equation} \label{eq:energy_3d}
    \Eddd=\int_{\Omddd}\hat{\mathcal{W}}(\bfF_e(s_1,s_2,s_3))
\end{equation}
We begin by passing to the limit as $h\to0$ to obtain a 2-d shell model. 
As is typical in methods of dimensionality reduction, in order to avoid trivial results -- i.e., either zero or infinite energy -- we first re-scale the energy by dividing by the small thickness parameter, $h$, and introduce the new variable $s_3'$ such that $s_3=hs_3'$. 
This allows us to write the energy density in a fixed, $h$-independent, re-scaled reference configuration occupying the space, $\Omdd\times[-1/2,1/2]$, where $\Omdd:=[0,L]\times[-a/2,a/2]$.
\begin{equation} \label{eq:enDen_rescale}
    \mathcal{E}_h=\frac{1}{h}\Eddd=\int_{\Omdd\times[-\half,\half]}\hat{\mathcal{W}}\left(\bfF_e(s_1,s_2,hs_3')\right)
\end{equation}
Under this rescaling, $\tilde{\bfu}$ and $\bfR$ remain the same, but $\bfxloc$ and $\nabla\bfxloc$ are transformed to:
\begin{equation} \label{posns_rescale}
    \bfxloc=\left\{\begin{array}{c} \chi\\ \yloc-\zloc_{,2}hs_3'\\ \zloc+\yloc_{,2}hs_3'\end{array}\right\}\quad,\quad \nabla\bfxloc=\begin{bmatrix}\chi_{,1} & \chi_{,2} & 0\\ \yloc_{,1}-\zloc_{,21}hs_3' & \yloc_{,2}-\zloc_{,22}hs_3' & -\zloc_{,2}\\ \zloc_{,1}+\yloc_{,21}hs_3' & \zloc_{,2}+\yloc_{,22}hs_3' & \yloc_{,2}\end{bmatrix}
\end{equation}
Furthermore, $\mrtheta$ (and by extension $\mrbfn$ and $\bfn$) become independent of $h$.
\begin{equation} \label{eq:theta_rescale}
    \mrtheta(s_3')=\left(\thetaTop+\frac{\pi}{4}\right)-\frac{\pi}{2}s_3'
\end{equation}

Lastly, in order to obtain finite -- i.e., non-zero and non-infinite -- curvatures in the limit as $h\to0$, it is necessary to scale the expansion coefficient, $\alpha$, linearly in $h$:
\begin{equation} \label{eq:alpha_rescale}
    \alpha=1+\frac{\alpha_0}{h_0}h,
\end{equation}
where $\alpha_0<0$ is a dimensionless parameter and $h/h_0\to 1$ as we take $h\to 0$. Note that the authors in \cite{agostiniani2017dimension} instead use $\alpha_0>0$ because they use the high-temperature state as their reference, where we use the low-temperature state.

Next, we use a linear series expansion of the re-scaled energy density about $h=0$ then integrate with respect to $s_3'$. Keeping in mind our definition of $\alpha$ from \eqref{eq:alpha_rescale}, the first term in \eqref{eq:enDen_index} has the following expansion:
\begin{equation}
    \alpha^{1/3} F^*_{ij}F^*_{ij}=\left[F^*_{ij}F^*_{ij}\right]_{h=0}+h\frac{\alpha_0}{3h_0}\left[F^*_{ij}F^*_{ij}\right]_{h=0}+2h\left[F^*_{ij}\right]_{h=0}\frac{\partial F^*_{ij}}{\partial h}+O(h^2)
\end{equation}
We note that the third term in the above expansion is linear in $s_3'$, and will therefore integrate to zero over $s_3'\in[-1/2,1/2]$. The remaining terms from the expansion are constant with respect to $s_3'$ and will therefore carry through to the areal strain energy unchanged. 

The second term in \eqref{eq:enDen_index} expands as follows:
\begin{equation}
    \left(\alpha^{-2/3}-\alpha^{1/3}\right)F_{ij}F_{kj}\mrn_i\mrn_k=-h\frac{\alpha_0}{h_0}\left[F_{ij}F_{kj}\mrn_i\mrn_k\right]_{h=0}+O(h^2)
\end{equation}
The only dependency on $s_3'$ in this expansion is in $n_in_k$ so, proceeding without using a specific form for the variation of $\bfn$ with $s'_3$, we have:
\begin{equation}
    \int_{-\half}^{\half}\left(\alpha^{-2/3}-\alpha^{1/3}\right)F_{ij}F_{kj}\mrn_i\mrn_k\,\dm s_3'=-\alpha_0\left[F_{ij}F_{kj}\right]_{h=0}\M[\mrn_i\mrn_k]+O(h^2)
\end{equation}
where we define the through-the-thickness averaging operator functional:
\begin{equation} \label{eq:averaging_operator}
    \M[\bullet]=\int_{-\half}^{\half}(\bullet)\dm s_3'
\end{equation}

For the volumetric terms, i.e., those that depend on $J$, we follow a different approach. 
Because the quantity $J$ has a clear geometric interpretation that we aim to preserve, we do not directly work with the volumetric energy density; we instead reduce $J$ to $J^h$, and then apply the energy to $J^h$.
Roughly, our approach is based on the recognition that the volumetric term $\hat{\mathcal{W}}$ is a penalty that approximates the incompressibility constraint $J=1$; therefore, we first identify $J^h$, and then apply the penalty $\hat{\mathcal{W}}$ to approximately enforce the constraint on $J^h$.

To obtain $J^h$, we first note that after re-scaling by the change of variables $s_3=hs_3'$), simply setting $h=0$ would make $J=0$. 
The appropriate quantity is $J/h$. 
We find the limit as $h\to0$ of $J/h$ and integrate with respect to $s_3'$:
\begin{equation} \label{eq:Jh_defn}
    J^h=\M\left[\lim_{h\to0}\frac{J}{h}\right]
\end{equation}
The aim of this rescaling is to avoid changing the topological structure of this penalty term in the integrand of the reduced functional. 
Maintaining this topological structure throughout the dimension reduction procedure also maintains the physical meaning of enforcing incompressibility through a penalty term. 
This heuristic idea works out, since the resulting expression satisfies $J^h = \det \bfF^h$ (defined below in \eqref{eq:defGrad_2d}), which mimics the behavior of the full 3-d description and shows that the reduction of $J$ to $J^h$ maintains the geometric meaning of the compressibility.

With the results above, we can take the limit as $h\to0$ to obtain a new reduced strain energy density function, $\mathcal{W}^h(s_1,s_2)$, given by:
\begin{equation} \label{eq:enDen_2d}
    \hat{\mathcal{W}}^h=\frac{\mu}{2}\left(\left(1+\frac{\alpha_0}{3}\right)\left(F^{*h}_{ij}F^{*h}_{ij}\right)-\alpha_0\left(F^h_{ji}F^h_{jk}\M\left[\mrn_i\mrn_k\right]\right)-3-2\log J^h\right)+\frac{\kappa}{2}\left(\Wvol\left(J^h\right)\right)
\end{equation}
where $\bfF^{h}(s_1,s_2)$ and $\bfF^{*h}(s_1,s_2)$ are the reduced deformation gradients given by:
\begin{equation} \label{eq:defGrad_2d}
    \bfF^h=\left.\bfF\right|_{s_3=0}=\left(\nabla\tilde{\bfu}+\left(\nabla\bfR^T\right)\bfxloch+\bfR^T\left(\nabla\bfxloc\right)^h\right)
\end{equation}
and
\begin{equation} \label{eq:f*h}
    \bfF^{*h}=\left.\bfF^*\right|_{s_3=0}=\bfF^h\left(\bfF_{eq}^{h}\right)^{-1}
\end{equation}
where the quantities that appear above are given by:
\begin{equation} \label{posn_2d}
    \bfxloch=\left\{\begin{array}{c} \chi\\ \yloc\\ \zloc\end{array}\right\}
    \implies
    \left(\nabla\bfxloc\right)^h=
    \begin{bmatrix}
        \chi_{,1} & \chi_{,2} & 0 \\
        \yloc_{,1} & \yloc_{,2} & -\zloc_{,2}\\
        \zloc_{,1} & \zloc_{,2} & \yloc_{,2}
    \end{bmatrix}
\end{equation}
and
\begin{equation} \label{eq:feq_reduced}
    \left(\bfF_{eq}^{h}\right)^{-1}=
    \begin{bmatrix}
        1 & 0 & 0\\
        0 & \cos{\left(2\mrbe s_2/a\right)} & \sin{\left(2\mrbe s_2/a\right)}\\
        0 & -\sin{\left(2\mrbe s_2/a\right)} & \cos{\left(2\mrbe s_2/a\right)}
    \end{bmatrix}
\end{equation}

Our 2-d energy functional is then given by
\begin{equation} \label{eq:energy_2d}
    \Edd=\int_{\Omdd}\mathcal{W}^h(s_1,s_2)
\end{equation}

We follow a slightly different procedure to pass to the 1-d model for LCE strips. We still rescale the energy by dividing by the width, $a$, and introduce the new coordinate $s_2'$ such that $s_2=as_2'$.
\begin{equation}
    \mathcal{E}^h_a=\frac{1}{a}\Edd=\int_{[0,L]\times[-\half,\half]}\mathcal{W}^h\left(s_1,as_2'\right)
\end{equation}
However, we now would like to retain some information about the width of the strip in an attempt to capture the difference in twisting behavior associated with \textit{wide} vs \textit{narrow} strips, defined by the value of the aspect ratio $a/L$. 
In \cite{seedPod}, they balanced bending and stretching energies to define a non-dimensional width parameter that depends on the width, length, and a so-called \textit{reference curvature} which measures the natural curvature associated with saddle-like equilibrium configurations. 
Though we have eliminated the dependence on the thickness in the previous steps, our reference curvature (i.e., the curvature associated with nematic activation in twisted-nematic configurations) is governed by $\alpha$, which we scaled with the thickness of the strip in \eqref{eq:alpha_rescale}.
Therefore, keeping the width of our strip, $a$, in the reduced 1-d energy density allows us to capture the behavior exhibited in \cite{seedPod}. 
So, to obtain the 1-d energy density, $\mathcal{W}^h_a$ without taking the limit as $a\to0$, we simply integrate the re-scaled 2-d energy density above with respect to $s_2'$. With the current construction, all terms are integrable in closed form in $s_2'$; however, since this final 1-d energy density does not have a clean representation, we simply use an overbar to represent integration with respect to $s_2'$ as follows:
\begin{equation} \label{eq:s2_int}
    \overline{(\cdot)}=\int_{-\half}^{\half}(\cdot)\dm s_2'
\end{equation}

The final lineic 1-d strain energy density is given by:
\begin{equation} \label{eq:enDen_1d}
    \hat{\mathcal{W}}^h_a=\frac{\mu}{2}\left(\left(1+\frac{\alpha_0}{3}\right)\left(\overline{F^{*h}_{ij}F^{*h}_{ij}}\right)-\alpha_0\left(\overline{F^h_{ji}F^h_{jk}\M\left[\mrn_i\mrn_k\right]}\right)-3-2\overline{\log J^h}\right)+\frac{\kappa}{2}\left(\overline{\Wvol\left(J^h\right)}\right)
\end{equation}
This reduced energy, together with the definitions of $\bfF^h$ and $\bfF^{*h}$ (from \eqref{eq:defGrad_2d} and \eqref{eq:f*h}, respectively) and the kinematic descriptions of $\yloc$, $\zloc$, and $\chi$ (from \eqref{eq:yloc_defn}, \eqref{eq:zloc_defn}, and \eqref{eq:xw_defn}, respectively) makes up our reduced order model. 
Tracing back these equations, the final energy density has first-order derivatives in the quantities $\bfu$, $\bfR$, and $\beta_e$.

%%%%%%%%%%%%%%%%%%%
%%%%%%%%%%%%%%%%%%%
%%%%%%%%%%%%%%%%%%%
%%%%%%%%%%%%%%%%%%%
\subsection{Description of Reduced Order Terms}

It should first be noted that this reduced form clearly separates the classical hyper-elasticity model and the effects of stimuli on the nematic liquid crystals. When $\alpha_0=0$, corresponding to zero external stimuli, the nematic terms vanish leaving us with a reduced compressible hyper-elasticity model where the effects of the changing cross-sectional shape are integrated into $\bfF^h$. To gain a better understanding of this reduced energy, we recall the three terms in $\bfF^h$.
\begin{enumerate}
    \item $\nabla\tilde{\bfu}$ captures the stretching and shear in the ribbon.
    \item $\left(\nabla\bfR^T\right)\bfxloch$ includes a gradient of the cross-section rotation, so it captures bending and twisting.
    \item $\bfR^T\left(\nabla\bfxloc\right)^h$ captures contributions from the changing cross-section shape through the gradient on $\bfxloc$.
\end{enumerate}
When $\bfF^h$ is contracted with itself in the energy density, there arises couplings between these different behaviors. 
For example, the first term contracted with itself gives the stretching energy, while the first term contracts with the second term to give the bending-stretching coupling.

For appropriate $\alpha_0<0$, by \eqref{eq:alpha_rescale} we have $0<\alpha<1$ since both $h$ and $h_0$ are positive. From \eqref{eq:F_l_defn}, it is clear that this would result in a compression along the nematic director and expansion transverse to the director, which is exactly the deformation associated with exposing a nematic LCE to external stimuli. In the reduced model, $\alpha_0<0$ weights the energy density differently, giving less weight to the purely geometric terms--which look like a reduced $\trace(\bfF^T \bfF)$ -- and more weight to the nematic terms -- which look like a reduced $|\bfF\bfn|^2$. Further, with no nematic activation (i.e., $\alpha_0=0$), the nematic terms are eliminated, leaving us with a form that looks like a reduced model for compressible Neo-Hookean materials.

%%%%%%%%%%%%%%%%%%%
%%%%%%%%%%%%%%%%%%%
\subsection{Boundary Conditions}
\label{sec:BCs}

Dirichlet boundary conditions are easily applied in the traditional sense for each of the reduced kinematic variables. 
However, the corresponding Neumann boundary conditions do not have an easy geometric or physical interpretation.

To illustrate this point, consider the traction boundary at the end of the LCE strip.
The normal vector at $s_1=L$ is approximately given by $\eOneLoc(L)$, so to apply the traction, $\bft$, at the end of the strip from the fully 3-d perspective, we would need to solve with the boundary constraints from the classical theory, given by:
\begin{equation} \label{eq: tBCs_center}
\begin{split}
    \bft &= J^{-1}\parderiv{\hat{W}}{\bfF}\bfF^T\eOneLoc\\
    &= J^{-1}\left(\mu\left(\bfF\-\bfF^{-T}\right)+\kappa\left(J-J^{-1}\right)J\bfF^{-T}\right)\bfF^T\eOneLoc\\
    &= \left(\mu J^{-1}\left(\bfF\bfF^T-\bfI\right)+\kappa\left(J-J^{-1}\right)\bfI\right)\eOneLoc
\end{split}
\end{equation}
To arrive at the appropriate reduced traction at the boundary, we should proceed by averaging the contributions to \eqref{eq: tBCs_center} over the entire cross-section of the strip by integrating over $s_2$ and $s_3$ to obtain a point constraint to be applied at $s_1=L$. This would produce a set of algebraic constraints on our reduced kinematic variables that would together describe the reduced boundary conditions.
However, this becomes algebraically formidable to compute due to the amount of inverses and matrix multiplications on the deformation gradient, and the resulting constraints on the reduced variables would be difficult to implement in non-trivial cases.
For this reason, the present work focuses only on calculations with Dirichlet and Neumann boundary conditions -- despite the unclear interpretation of the latter -- on the reduced kinematic variables.

%%%%%%%%%%%%%%%%%%%
%%%%%%%%%%%%%%%%%%%
\section{Numerical Method}
\label{sec:num-mthd}

%%%%%%%%%%%%%%%%%%%
%%%%%%%%%%%%%%%%%%%
%%%%%%%%%%%%%%%%%%%
%%%%%%%%%%%%%%%%%%%
\subsection{Numerical Treatment of Finite Rotations}
\label{sec:quaternions}

To parameterize the 3-d finite rotations, we follow \cite{GUINOT201273} in using unit quaternions for their noted numerical stability and computational efficiency. The 3-d axial rotation vector associated with the rotation tensor, $\bfR(s_1)$, is represented by a 4D vector, $\bfq(s_1) = (q_0, q_1, q_2, q_3)$ using the following relationship:
\begin{equation}
\bfR=\begin{bmatrix}(1-2q_2^2-2q_3^2) & 2(q_1q_2-q_0q_3) & 2(q_1q_3+q_0q_2)\\2(q_1q_2+q_0q_3) & (1-2q_1^2-2q_3^2) & 2(q_2q_3-q_0q_1)\\2(q_1q_3-q_0q_2) & 2(q_2q_3+q_0q_1) & (1-2q_1^2-2q_2^2)\end{bmatrix}
\end{equation}
where $q_0(s_1)$, $q_1(s_1)$, $q_2(s_1)$, and $q_3(s_1)$ are dimensionless, and $\bfq(s_1)$ is a unit vector:
\begin{equation}
    q_0^2+q_1^2+q_2^2+q_3^2=1.
\end{equation}
These quaternions are related to the rotation tensor, $\bfR$, through its axial vector, $\bfr$, by the following:
\begin{equation}
    q_0=\cos{\frac{r}{2}}\quad,\quad q_1=v_1\sin{\frac{r}{2}}\quad,\quad q_2=v_2\sin{\frac{r}{2}}\quad,\quad q_3=v_3\sin{\frac{r}{2}}
\end{equation}
where $r=\|\bfr\|$ is the magnitude of the rotation and $\bfv=\bfr/r$ is the unit vector in the direction of the axis of rotation. The unit-length constraint on the quaternions will be implemented with Lagrange multipliers in the lineic strain energy density:
\begin{equation}
    \tilde{\mathcal{W}}^h_a=\hat{\mathcal{W}}^h_a+\lambda(q_0^2+q_1^2+q_2^2+q_3^2-1)
\end{equation}

%%%%%%%%%%%%%%%%%%%
%%%%%%%%%%%%%%%%%%%
%%%%%%%%%%%%%%%%%%%
%%%%%%%%%%%%%%%%%%%
\subsection{Finite Element Discretization}

For this work, we find the deformation by minimizing the potential energy (subject to appropriate boundary conditions) with respect to the following reduced kinematic variables:
\begin{itemize}
    \item cross-section displacement: $u_1,u_2,u_3$
    \item cross-section rotation: $q_0,q_1,q_2,q_3$
    \item cross-section opening angle: $\beta_e$
\end{itemize}
We minimize the energy in this nonlinear problem in the usual way: we start with an initial guess for the kinematic variables, evaluate the functional derivative, and update to move in the steepest descent direction.
At each step of the iteration, we calculate the current nematic director, $\bfn$, by reconstructing the full 3-d deformation gradient \eqref{eq:def_F} and mapping the nematic director as in \eqref{eq:director_map}, to provide the required expressions for the director field in the energy.

For the finite element discretization, we use standard 1-d finite elements with quadratic interpolation for the reduced kinematic variables, and linear interpolation for the Lagrange multiplier that constrains the quaternions to unit length.
This is in line with the heuristic for constrained problems of using a lower-order interpolation for the Lagrange multipliers.

%%%%%%%%%%%%%%%%%%%
%%%%%%%%%%%%%%%%%%%
%%%%%%%%%%%%%%%%%%%
%%%%%%%%%%%%%%%%%%%
\section{Numerical Results}
\label{sec:results}

In this section, we present some results for our 1-d model, computed numerically. 
We first present examples to validate the model.
Since the main goals are to capture (1) the spontaneous curvature behavior of twisted nematic LCEs, and (2) the classic behaviors of tape springs, we test the ability of the 1-d model to capture these effects. 
We first compute in Section \ref{sec:results-nem-active} the spontaneous curvature of twisted nematic LCEs under nematic activation at different offset angles using the 1-d model. 
Next, in Section \ref{sec:results-tapeSpring}, we present our results from the 1-d model for the tape-spring instability and twist-bend coupling of transversely curved strips. 
To validate our model, we compare with the results of a full 3-d model in Section \ref{sec:validation}. 
Finally, after validation, we present further results from our reduced model that we compare qualitatively to experiments from \cite{CLEMENT2021101362}.
We highlight that the comparisons to experiment are only qualitative because the experiments of \cite{CLEMENT2021101362} -- as well as most others in the literature -- do not quantitatively measure deformation, stress, or other quantities that can be used for a quantitative test of the model.

%%%%%%%%%%%%%%%%%%%
%%%%%%%%%%%%%%%%%%%
%%%%%%%%%%%%%%%%%%%
%%%%%%%%%%%%%%%%%%%
\subsection{Heating/Illumination in a Twisted Nematic LCE}
\label{sec:results-nem-active}

\begin{figure}[htb]
    \centering
    \includegraphics[width=10cm]{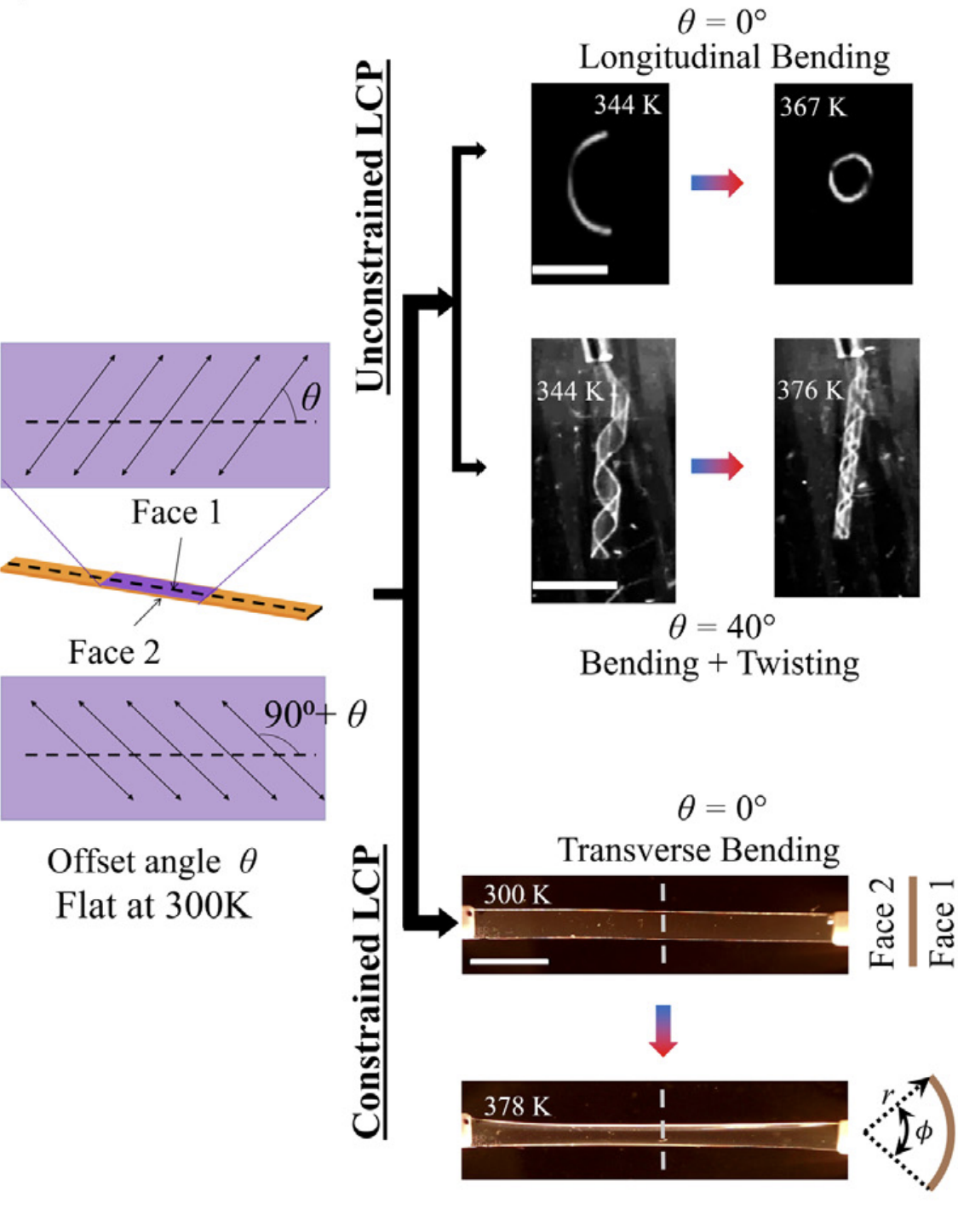}
    \caption{This figure from \cite{CLEMENT2021101362} shows experimental observations of heating a twisted nematic LCE strip with different cut angles, $\theta$. 
    Top right: for an unconstrained strip, $\theta=0^\circ$ shows pure bending, while $\theta=40^\circ$ forms a helical shape with combined bending and twisting. 
    Bottom right: by constraining the ends of the strip, we see that a spontaneous cross-sectional curvature develops from a strip with $\theta=0^\circ$.\label{fig:ravi}}
\end{figure}

\begin{figure}[ht!]
    \subfloat[]{\includegraphics[width=0.47\textwidth]{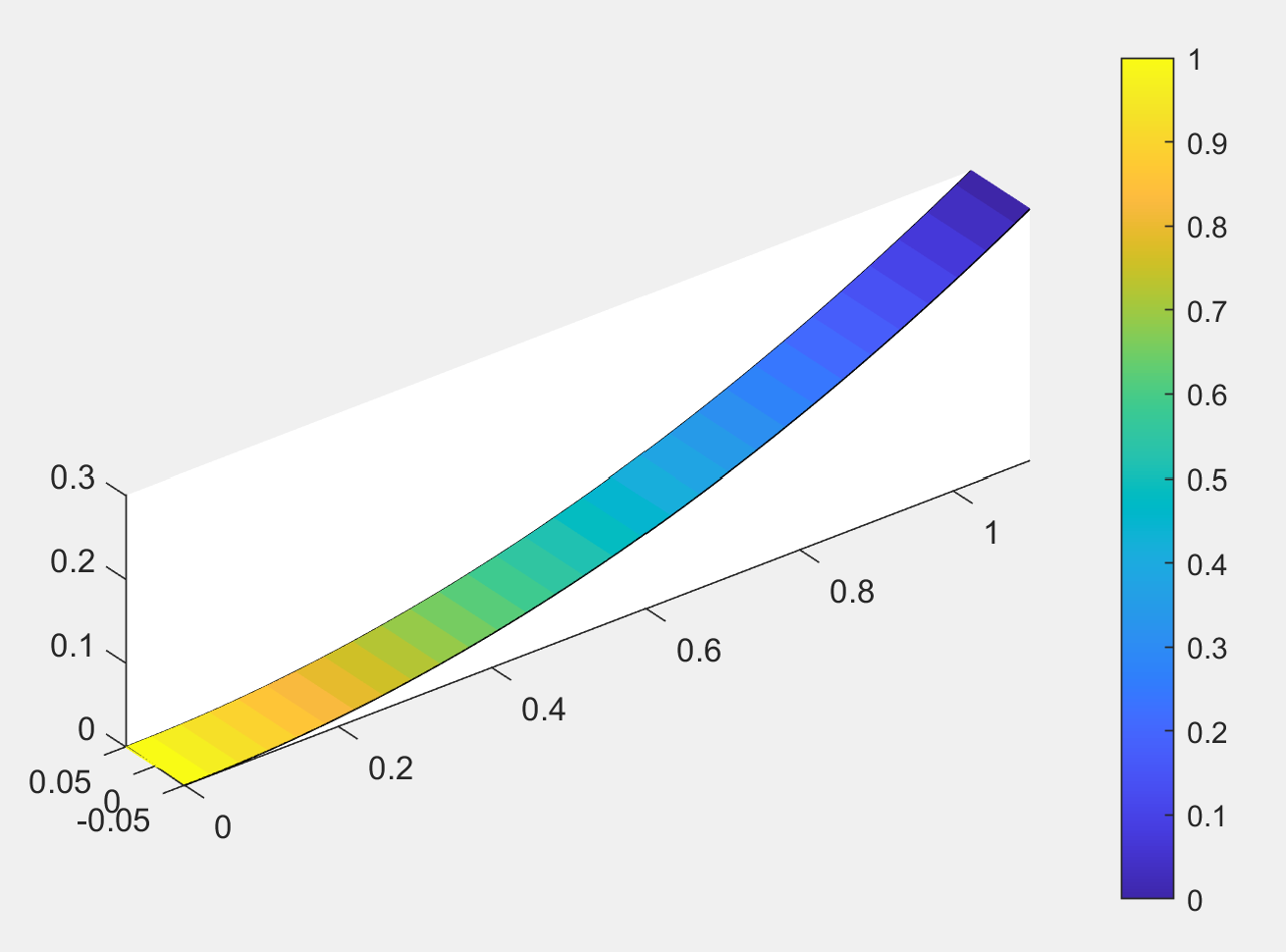}}
    \hfill
    \subfloat[]{\includegraphics[width=0.47\textwidth]{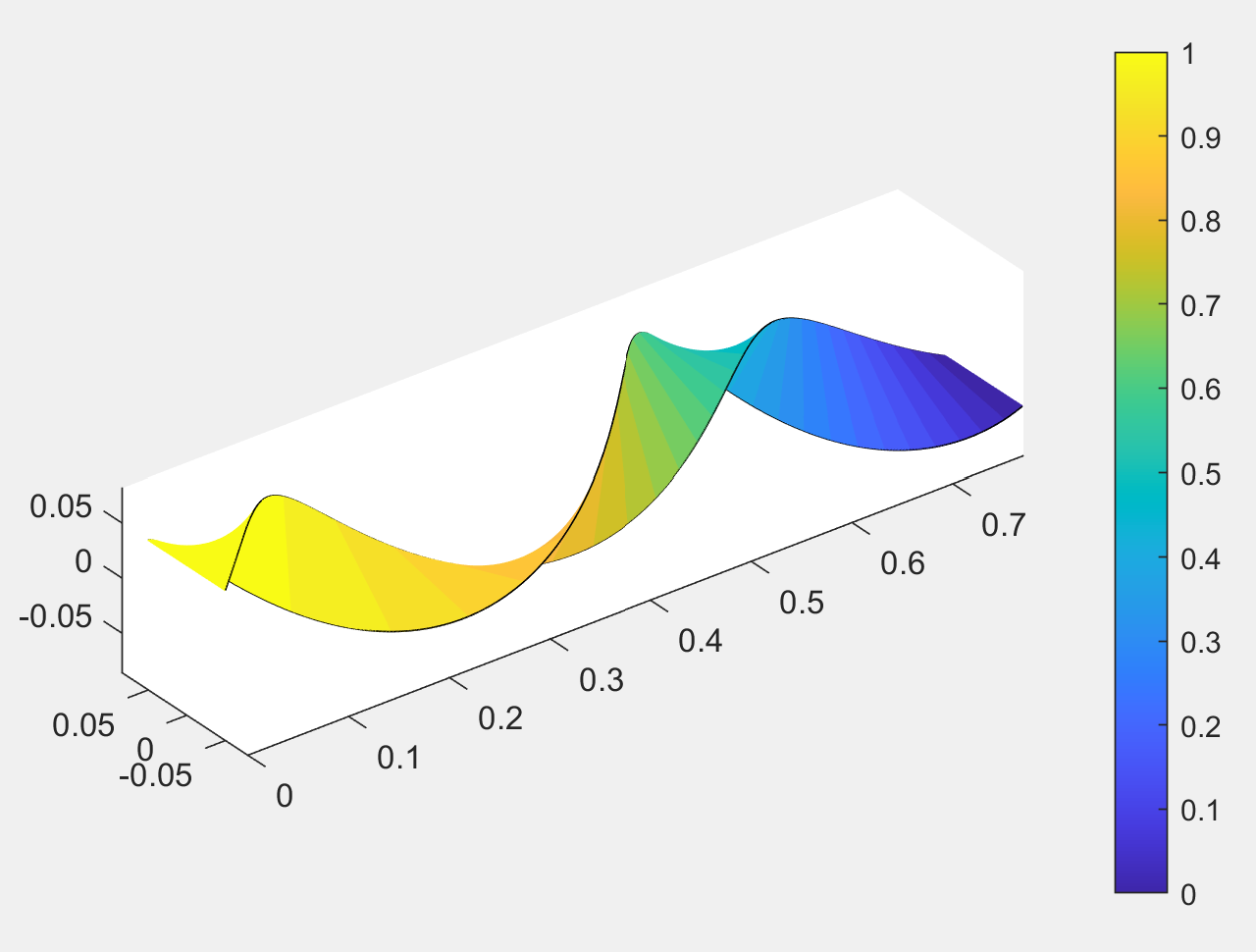}}
    \caption{(a) An initially-flat LCE strip with no nematic offset angle will undergo bending upon nematic activation. The expansion along the length on the bottom face and the opposing contraction on the top face create this distinct behavior. (b) For an initially-flat, twisted nematic LCE strip with an offset angle, $\theta_{top}=40^{\circ}$, we obtain a combination of bending and twisting upon nematic activation. 
    In both calculations, the left end of the strip was held fixed (Dirichlet boundary conditions) while the right end was free (Neumann boundary conditions). The color corresponds to the lineic strain energy density, normalized to be between $0$ and $1$.}
    \label{fig:LCE_bend}
\end{figure}

We capture the experimental observations shown in Figure \ref{fig:ravi} from \cite{CLEMENT2021101362} using the 1-d model; numerical results are shown in Figure \ref{fig:LCE_bend} with the colors corresponding to the normalized lineic strain energy density. 

Figure \ref{fig:LCE_bend} present results for two initially-flat LCE strips with a twisted nematic director configuration. 
The first has an offset angle, $\theta_{top}=0^{\circ}$, while the second has $\theta_{top}=40^{\circ}$. 
We hold the left end fixed with zero displacement ($u_1=u_2=u_3=0$), zero rotation ($q_0=1$, $q_1=q_2=q_3=0$), and zero transverse curvature ($\beta_e=0)$. 
After nematic activation, the strip with  $\theta_{top}=0^{\circ}$ bends upward as seen in the experiments in Figure \ref{fig:ravi}; this is because the strip expands along the length on the bottom side of the strip and contracts on the top side. 
For the strip with  $\theta_{top}=40^{\circ}$, the model captures the observed combined bending and twisting.

%%%%%%%%%%%%%%%%%%%
%%%%%%%%%%%%%%%%%%%
%%%%%%%%%%%%%%%%%%%
%%%%%%%%%%%%%%%%%%%
\subsection{Classical Results in the Analysis of Tape-Springs}
\label{sec:results-tapeSpring}

A key goal of this work is to be able to effectively capture the localizations in LCE strips that are closely related to the classical tape-spring instability observed in the purely mechanical setting in transversely curved strips.
Here, we show that the 1-d model is able to capture this instability, as well as well-known coupling between twisting and bending, in a purely mechanical setting.

\begin{figure}[htb!]
    \centering
    \includegraphics[width=10cm]{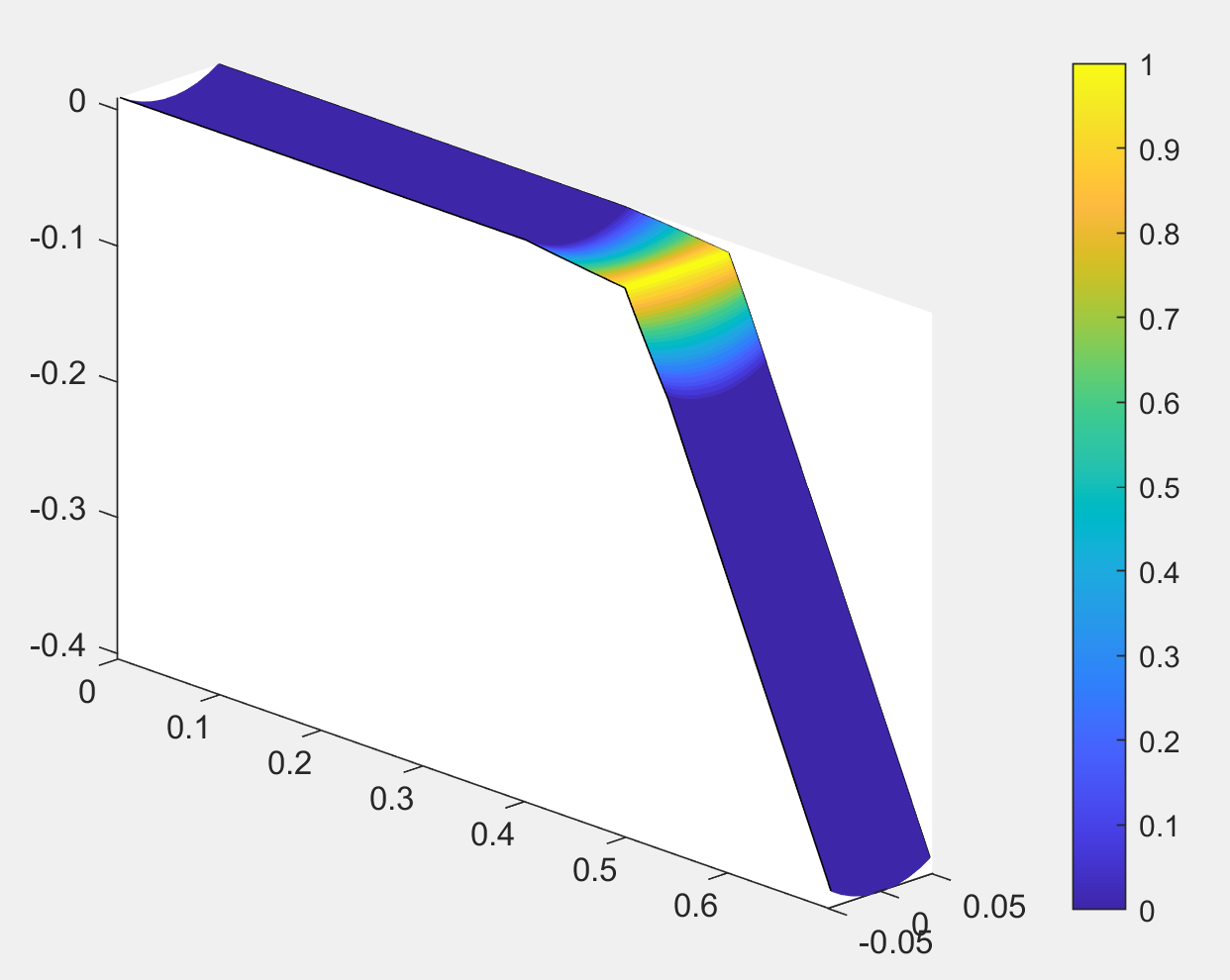}
    \caption{The 1-d reduced model captures classical tape-spring instability in pre-curved LCE strips under purely mechanical loading without nematic activation. 
    The color shows the difference in opening angle, $\beta_e$, between the natural and deformed configurations, normalized to be between 0 and 1, and highlights the flattening of the cross-section that goes with the localized bending. \label{fig:tapeSpring}}
\end{figure}

To model the tape-spring instability, the cross-section is set to have transverse curvature in its natural configuration. 
The left end of the strip is held fixed, while a downward force is applied to the right end. 
For ease of visualization, we also applied the equivalent of a roller boundary condition in the middle of the strip, i.e., $u_3(s_1=0.5)=0$. 
As shown in Figure \ref{fig:tapeSpring}, the bending (and strain energy) localizes in a small region, and the transverse curvature vanishes here. 
The color corresponds to the change in opening angle $\beta_e$ as compared to the natural configuration to highlight the flattening cross-section typical to the tape-spring instability. 
We note that our lineic energy density contains first-order derivatives of $\beta_e$ which, roughly, keeps it continuous, so the tape-spring localization does not become singular and can be captured with a standard finite element discretization.

\begin{figure}[htb!]
    \centering
    \includegraphics[width=10cm]{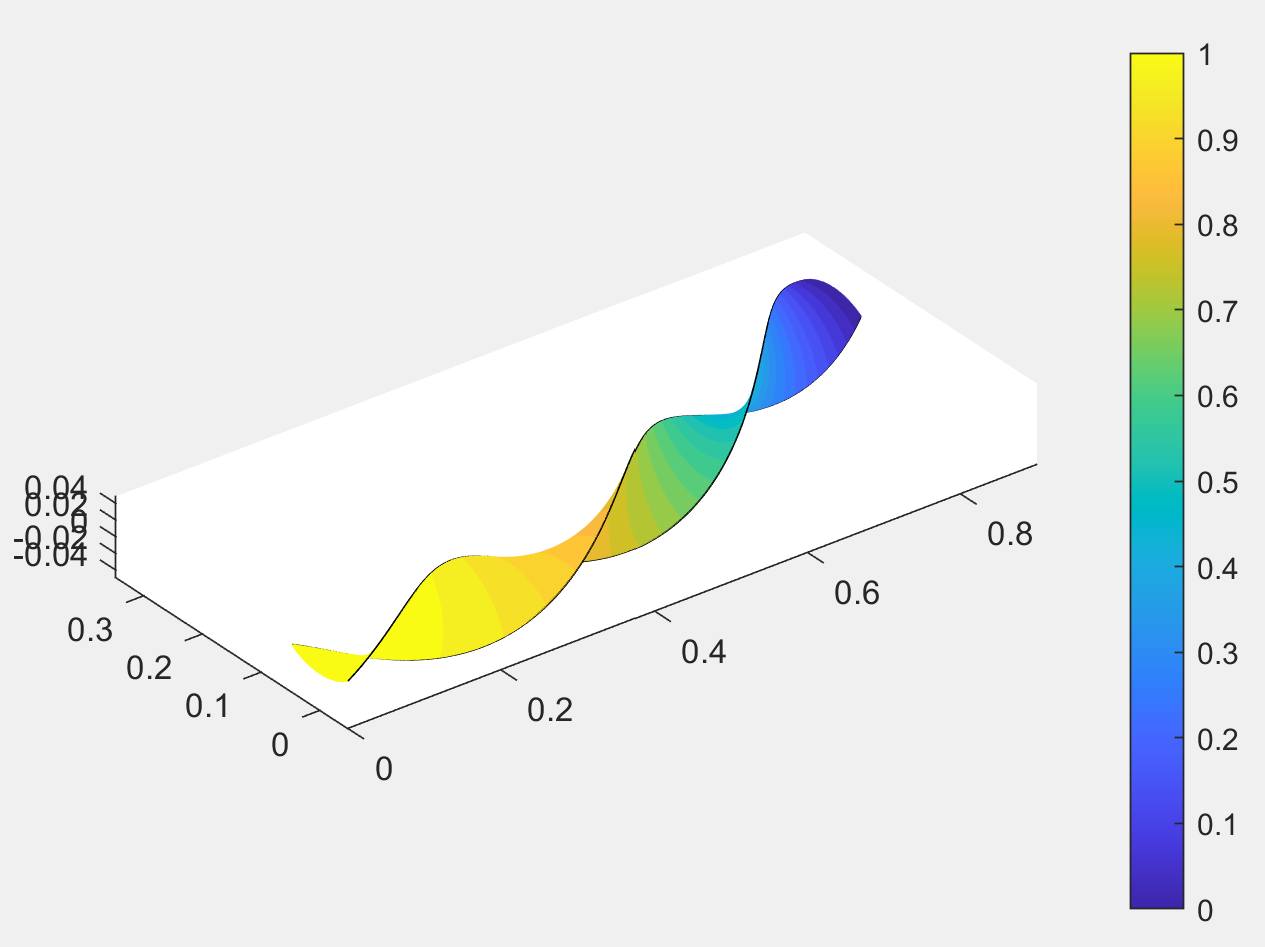}
    \caption{The non-linearity in the reduced model can capture the coupling between twisting and bending under a torsional load for transversely curved LCE strips without nematic activation. For this calculation, the left end of the strip is held fixed ($u_i(0)=q_i(0)=0$ for $i=1,2,3$; $q_0(0)=1$) while the right end of the strip is subjected to a rotation ($q_0(L)=q_2(L)=q_3(L)=0$; $q_1(L)=1$) while keeping displacements free. The uneven warping throughout the cross-section results in a same-sense bending that is coupled to the twisting.  The color corresponds to the lineic strain energy density, normalized to be between $0$ and $1$. \label{fig:tapeTwist}
    }
\end{figure}

Next, we demonstrate the ability of our model to capture coupled behavior. 
In particular, the cross-section was assigned a transverse curvature in the natural configuration, and a torque was applied to the right end while the left end was held fixed. Due to the cross-sectional curvature, the twisting was coupled with deformation in the same sense as the twisting, as shown in Figure \ref{fig:tapeTwist}.

%%%%%%%%%%%%%%%%%%%
%%%%%%%%%%%%%%%%%%%
%%%%%%%%%%%%%%%%%%%
%%%%%%%%%%%%%%%%%%%
\subsection{Model Validation}

\label{sec:validation}
\begin{figure}[ht!]
    \subfloat[]{\includegraphics[width=0.47\textwidth]{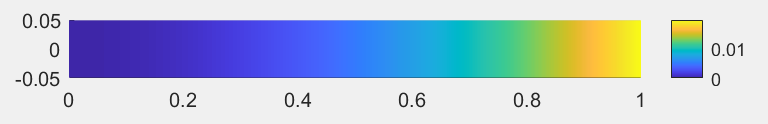}}
    \hfill
    \subfloat[]{\includegraphics[width=0.47\textwidth]{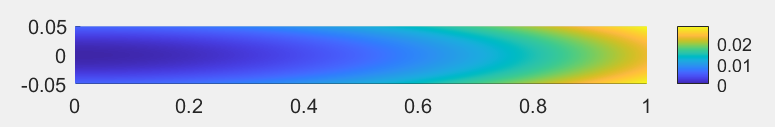}}
    \\
    \subfloat[]{\includegraphics[width=0.47\textwidth]{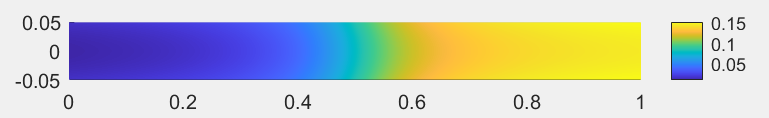}}
    \hfill
    \subfloat[]{\includegraphics[width=0.47\textwidth]{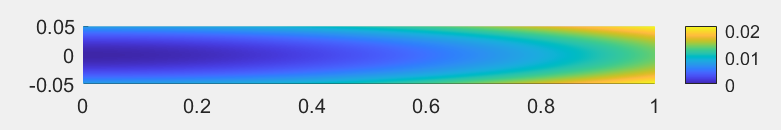}}
    \caption{We define the error as the absolute value of the percent difference between the displacement in the 3-d model and the 1-d model. We then average this error through the thickness of the strip to provide a clear picture of the efficacy of the reduced model. We show here this error for
    \\ (a) the bending of a strip via nematic activation (see Figure \ref{fig:LCE_bend}(a)); 
    \\ (b) the bending and twisting of a strip via nematic activation (see Figure \ref{fig:LCE_bend}(b));
    \\ (c) the tape-spring instability (see Figure \ref{fig:tapeSpring}); and 
    \\ (d) the coupled twisting and bending of a tape spring (see Figure \ref{fig:tapeTwist}).}
    \label{fig:error_plots}
\end{figure}

While our 1-d model qualitatively captures a large range of behaviors, here we quantitatively compare the accuracy against the 3-d model, noting that the experimental observations are not quantitative and hence do not enable a comparison between 1-d model and experiment.
Figure \ref{fig:error_plots} shows the relative error of our reduced model as compared to a full 3-d FEM simulation. 
The error is defined as the absolute value of the percent difference between the two solutions, and it is averaged through the thickness of the strip to provide an easy visualization of the efficacy of the reduced model. 
One trend that immediately emerges is that the reduced model performs worse as we move away from the centerline. 
This is an expected outcome; constraining the cross-section curve to remain open-circular and taking Taylor expansions for small opening angles, $\beta_e$, is going to have the largest effects far from the centerline. 
Further, this increase is expected to be more prominent in regions where we have rapidly-varying transverse curvature as can be seen in the middle of the strip for the tape-spring instability (Figure \ref{fig:error_plots}(c)) and the coupled bending and twisting of a tape-spring (Figure \ref{fig:error_plots}(d)).
Also to be expected, in the case where the strip is flat and only bends (Figure \ref{fig:error_plots}(a)), there is not appear to be any contribution to the error away from the centerline.

Another trend evident in the error plots is the increase in error toward the free end of the strip. 
This is most evident in the case of the tape-spring instability (Figure \ref{fig:error_plots}(c)), in which the error seems to increase drastically after the localization.
This is likely due to the error in the bend angle at the point of the localization, which then propagates because the rotations of the right portion of the tape-spring are consequently incorrect.
It could also be due to the fact that we do not use the carefully constructed reduced form for the traction boundary conditions described in Section \ref{sec:BCs}; instead, we have used the natural boundary conditions that arise from our reduced kinematic variables in the weak form of the reduced energy. 

%%%%%%%%%%%%%%%%%%%
%%%%%%%%%%%%%%%%%%%
%%%%%%%%%%%%%%%%%%%
%%%%%%%%%%%%%%%%%%%
\subsection{Development of Cross-Section Curvature in a Constrained LCE}
\label{sec:results-constrainedLCE}

In this section, we show the 1-d model can capture the well-known behavior of spontaneous curvature formation under nematic activation for a strip with the twisted-nematic director alignment; see the bottom of Figure \ref{fig:ravi} for the experimental observation.

\begin{figure}[htb!]
    \centering
    \includegraphics[width=10cm]{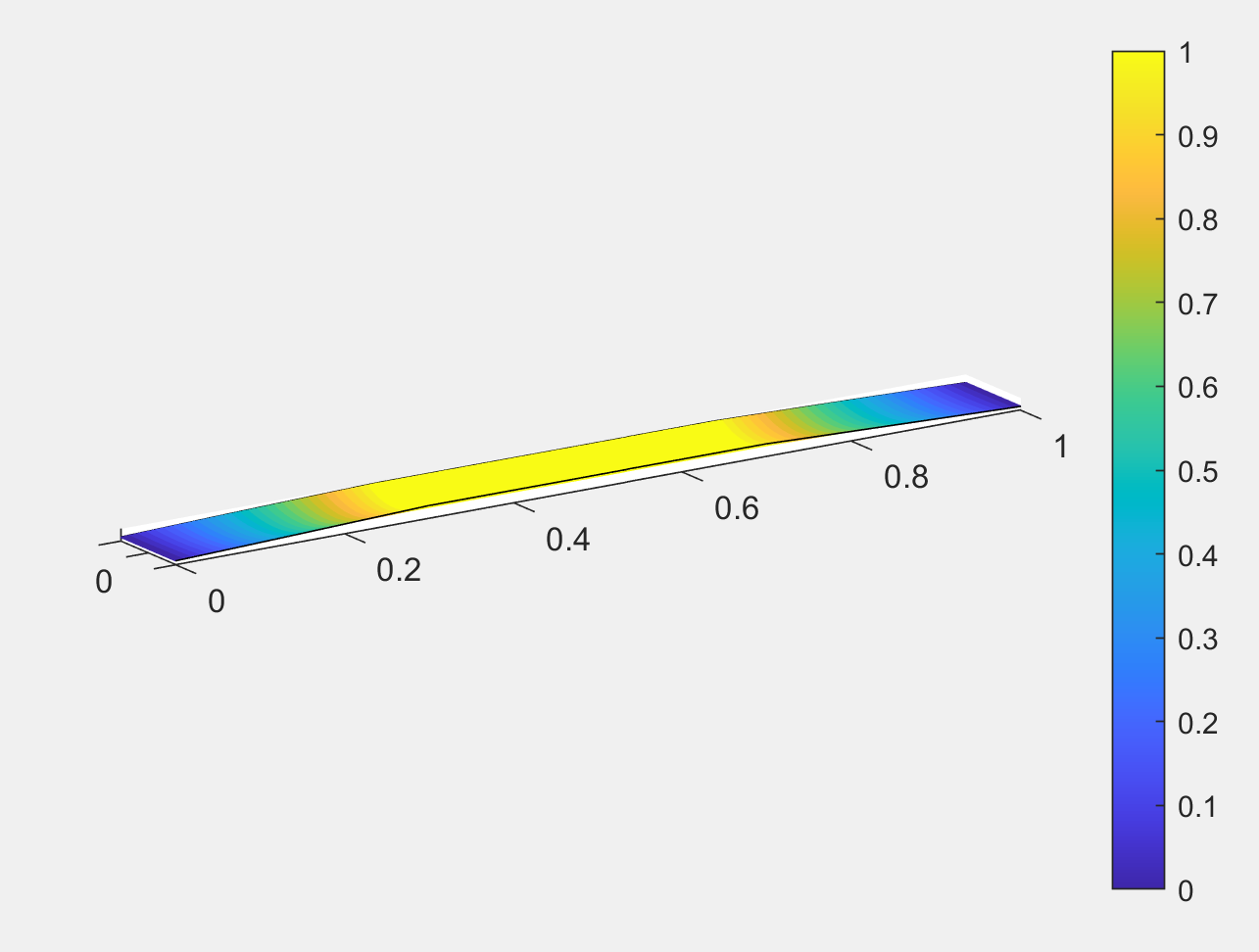}
    \caption{Nematic activation in a twisted nematic configuration causes spontaneous \textit{transverse} curvature in a constrained LCE strip. Held fixed at both ends, the curvature cannot be attained along the length of the strip,  so the minimum energy configuration attains transverse curvature along the width of the strip instead. This reproduces the experimental observation shown at the bottom of Figure \ref{fig:ravi}.
    The color shows the transverse curvature (given by the opening angle), normalized to be between 0 and 1.
    \label{fig:spontCurve}
    }
\end{figure}

In this case, the director is oriented as in Figure \ref{fig:twist-nem}.
This initially flat strip was completely fixed at both ends: zero displacement ($u_1=u_2=u_3=0$), zero rotation ($q_0=1$, $q_1=q_2=q_3=0$), and held with a flat cross-section ($\beta_e=0$). 
As shown in Figure \ref{fig:spontCurve}, the strip develops transverse curvature as it is prevented from bending along its length. 
In the transverse direction, the strip elongates on the bottom face while shrinking along on the top face. 

%%%%%%%%%%%%%%%%%%%
%%%%%%%%%%%%%%%%%%%
%%%%%%%%%%%%%%%%%%%
%%%%%%%%%%%%%%%%%%%
\subsection{Tape-Spring Localizations in Activated LCE}
\label{sec:results-3bar}

\begin{figure}[htb!]
    \subfloat[]{\includegraphics[width=0.4\textwidth]{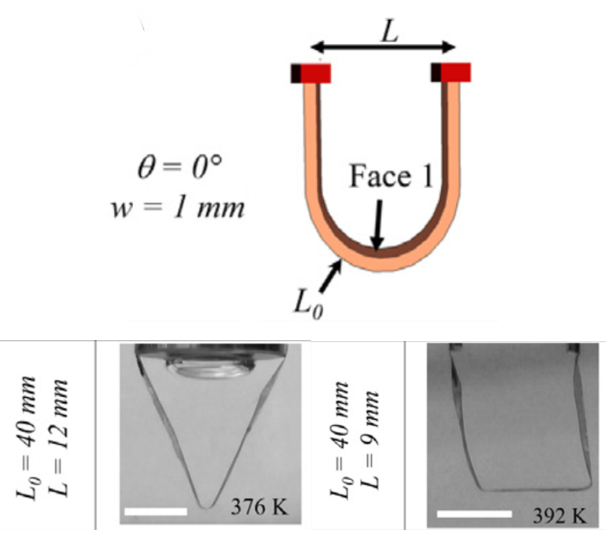}}
    \\
    \subfloat[]{\includegraphics[width=0.4\textwidth]{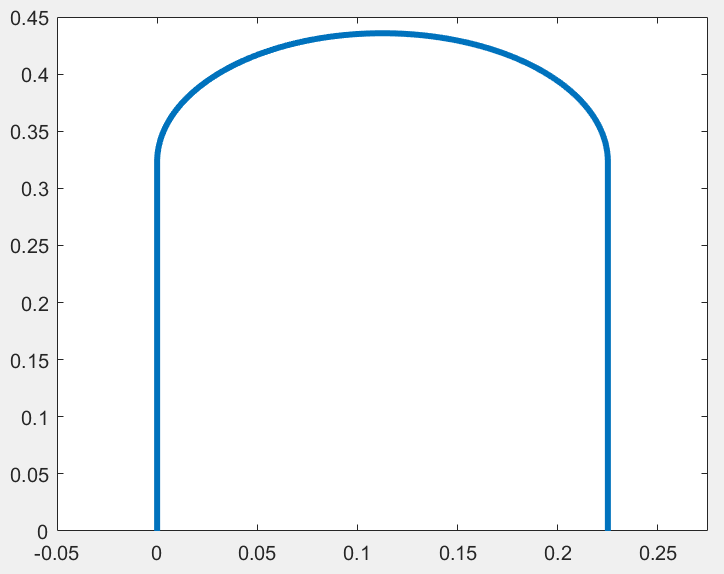}}
    \hfill
    \subfloat[]{\includegraphics[width=0.4\textwidth]{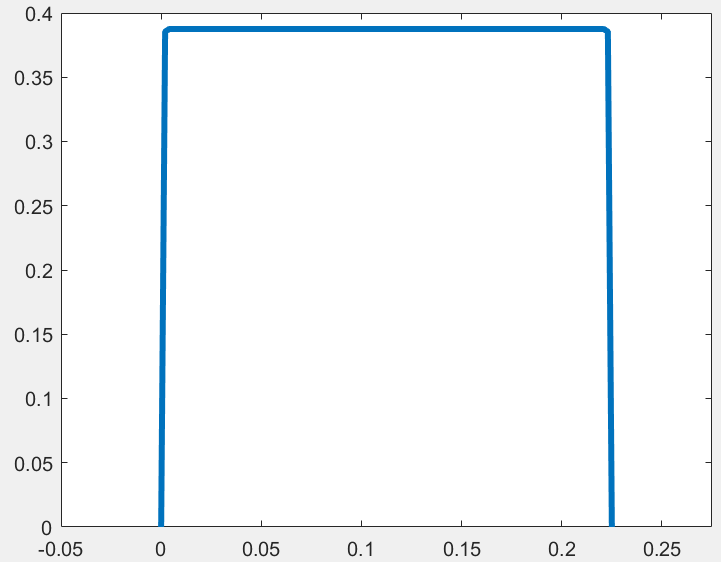}}
    \\
    \subfloat[]{\includegraphics[width=0.4\textwidth]{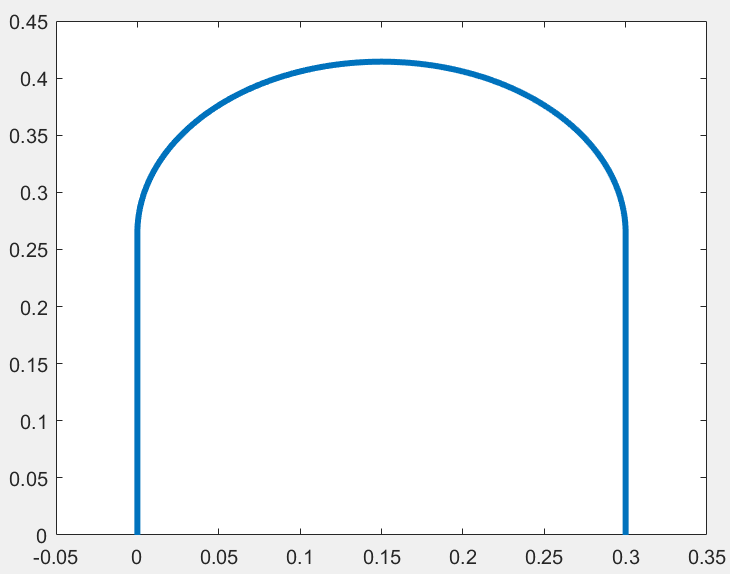}}
    \hfill
    \subfloat[]{\includegraphics[width=0.4\textwidth]{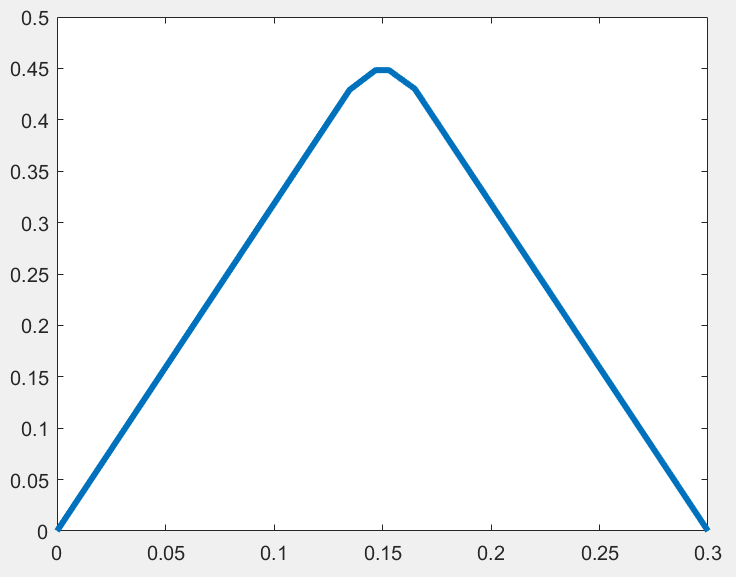}}
    \caption{
        The 1-d model captures the interactions between tape-spring instabilities due to transverse curvature and nematic activation, following the observations from \cite{CLEMENT2021101362} shown in (a).
        Here, we see results of an initially flat strip with $\theta_{top}=0$ under different constrained loading conditions, where we have plotted the centerline position.
        In (b) and (d), we have the purely mechanical behavior of the bent strip prior to nematic activation, with $u_1(L) < 0$ and $u_1(0)=u_2(0)=u_3(0)=u_2(L)=u_3(L)=0$. 
        In (c), the curvature in (b) localizes in two different locations upon nematic activation. 
        In (e), the curvature in (d) localizes in only one location upon nematic activation. 
        The only difference between these calculations is the value prescribed to $u_1(L)$: in (b) we use $u_1(L)=-0.775$, and in (d) we use $u_1(L)=-0.7$.
    } 
    \label{fig:mechanisms}
\end{figure}

We next examine intriguing experiments from \cite{CLEMENT2021101362} in which there is an interplay between the tape-spring localization behavior and the nematic activation.
In the experiments shown in Figure \ref{fig:mechanisms}(e), initially-flat LCE strips are bent into a U-shape.
The constrained boundary conditions lead to the formation of transverse curvature, as discussed in Section \ref{sec:results-constrainedLCE}. 
However, the transverse curvature interacts with the U-shape to localize much like the classical tape-spring instability.
Depending on how far apart the ends are held fixed, these strips develop either 1 or 2 localizations upon nematic activation: with the ends held farther apart, we get a single localization, but when the ends are held closer together, we can obtain two localizations.

Figure  \ref{fig:mechanisms} shows the predictions from the 1-d model that captures this behavior.

%%%%%%%%%%%%%%%%%%%
%%%%%%%%%%%%%%%%%%%
%%%%%%%%%%%%%%%%%%%
%%%%%%%%%%%%%%%%%%%
\subsection{Effect of Aspect Ratio: Wide v. Narrow ribbons}

As noted previously, the width of the strip is a parameter in our 1-d model, allowing us to differentiate between strips of different width. 
As described in \cite{seedPod}, narrow ribbons display pure twisting, while wider ribbons display a combination of bending and twisting. 
All of the results we presented in the previous examples used a non-dimensional width of $a/L=0.1$, and here we examine the effect of using a narrower ribbon with $a/L=0.05$.

\begin{figure}[htb!]
    \centering
    \includegraphics[width=10cm]{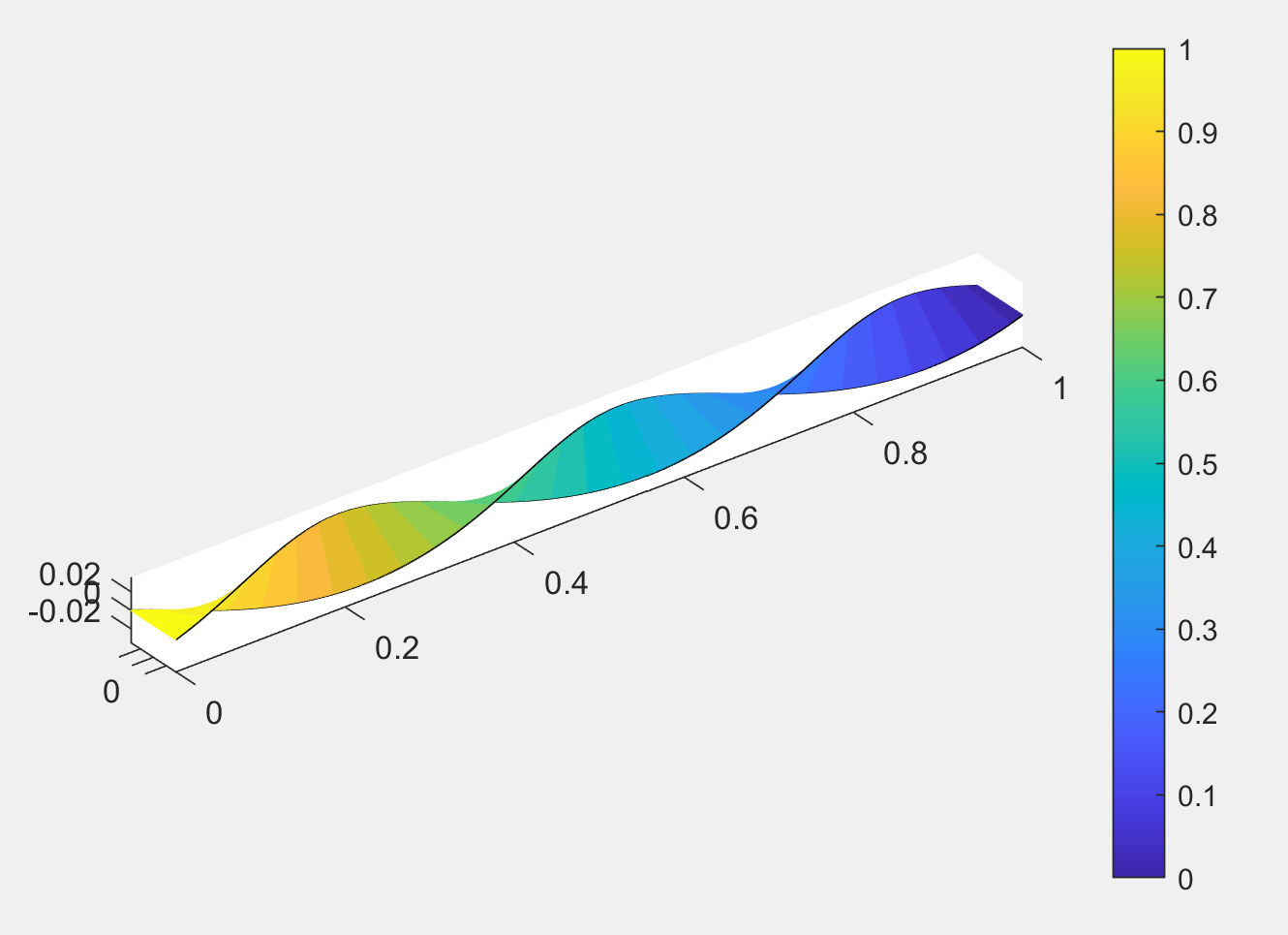}
    \caption{As opposed to the \textit{wide} ribbon shown in Figure \ref{fig:LCE_bend}(b) that exhibited a combination of bending and twisting, this \textit{narrow} ribbon exhibits pure twisting.  The color corresponds to the lineic strain energy density, normalized to be between $0$ and $1$. \label{fig:narrow ribbons}
    }
\end{figure}

Figure \ref{fig:narrow ribbons} compares the effect of aspect ratio by examining an LCE strip that has all parameters and boundary conditions as in Figure \ref{fig:LCE_bend}(b), with the exception that the former has aspect ration $a/L=0.05$ (narrow) and the latter has $a/L = 0.1$ (wide).
We see that the narrow ribbon in Figure \ref{fig:narrow ribbons} shows pure twisting, while the wide ribbon in Figure \ref{fig:LCE_bend}(b) showed a combination of bending and twisting, broadly in accord with \cite{seedPod}.

%%%%%%%%%%%%%%%%%%%
%%%%%%%%%%%%%%%%%%%
%%%%%%%%%%%%%%%%%%%
%%%%%%%%%%%%%%%%%%%
\section{Discussion}
\label{sec:discussion}

We have presented a 1-d model for transversely-curved LCE strips.
Our work was motivated by the experiments of \cite{CLEMENT2021101362} which show that the transverse curvature leads to a complex set of configurations upon nematic activation.
We show that the model captures well a large range of observed behavior, including localization instabilities that are related to the classical tape-spring instability.
The 1-d nature of the final dimensionally-reduced model enables rapid exploration of the design space, and the match to experimental results provides confidence in the model being suitable for this purpose.

While the success of the 1-d model is promising and useful to realistically model experiments, the following directions for future efforts appear to be promising:
\begin{enumerate}

    \item generalization to non-circular cross-sections: A simplifying assumption in this work was that the deformed strips have open circular cross-sections. It is straightforward to relax this assumption, at the expense of (significantly) more complex algebraic expressions. However, this can potentially significantly improve the accuracy in capturing the tape-spring localizations.

    \item consistent dimensionally-reduced boundary conditions: As discussed in Section \ref{sec:BCs}, the physical meaning of applying Neumann boundary conditions is not transparent, while the consistent dimensionally-reduced boundary conditions can be algebraically formidable and potentially nonlinearly couple between different kinematic degrees of freedom. Applying these consistent boundary conditions as constraints may provide a feasible path forward, and could potentially be important to accurately capturing the tape-spring localization.

    \item loops with overcurvature and self-contact: The experiments in \cite{CLEMENT2021101362} and analysis in \cite{purohit2008plectoneme,coleman2000theory,van2003instability} show that overcurvature of closed loops leads to complex and interesting configurations, including self-contact.
    This offers a rich design space that would be valuable to explore with an augmentation of our model that accounts for self-contact.
 
     \item rigorous dimension reduction based on $\Gamma$-convergence: The ansatz-based approach in this work could be made more rigorous using the mathematical framework of $\Gamma$-convergence, following several prior results including the seminal work of Friesecke, James, and Muller \cite{friesecke2006hierarchy} as well as others in the context of LCE \cite{agostiniani2011gamma} and Friesecke-James-Muller.
    The ansatz proposed in this work provides a natural starting point to construct the recovery sequence for such a $\Gamma$-convergence analysis.
    
    \item methods of dimension-reduction drawing from data science: Recent methods in data science aimed at reducing the dimensionality of datasets, such as diffusion maps \cite{coifman2006diffusion}, can potentially provide insights into the appropriate variables and ansatz for dimension reduction by treating the numerical solution of the 3-d problem as high-dimensional ``data''.

    \item the comparison to experiment is largely qualitative, and is due to the fact that quantitative experimental measurements of deformation and shape are very challenging and a focus of current activity in several research groups. When richer experimental measurements are available in the future, it will be essential to test our model against these observations.

\end{enumerate}

%%%%%%%%%%%%%%%%%%%%% 
%%%%%%%%%%%%%%%%%%%%%
%%%%%%%%%%%%%%%%%%%%%
%%%%%%%%%%%%%%%%%%%%%
\section*{Software Availability}

The symbolic algebra calculations (in Mathematica) and the numerical implementation (in Matlab) are available at this link:
\url{github.com/klogrande/LCE}

\section*{Conflicts of interest}

There are no conflicts of interest to declare.

\begin{acknowledgments}
    We thank Mahnoush Babaei and Noel Walkington for useful discussions; 
    NSF XSEDE for computing resources provided by the Pittsburgh Supercomputing Center;
    and the DOD SMART Fellowship program, NSF (MOMS 1635407, DMREF 1921857, DMS 2108784), ARO (W911NF-17-1-0084), BSF (2018183), and AFOSR (MURI FA9550-18-1-0095) for financial support.
    Figures \ref{fig:ravi} and \ref{fig:mechanisms}(a) are reprinted from \cite{CLEMENT2021101362} with permission from Elsevier.
\end{acknowledgments}

%%%%%%%%%%%%%%%%%%%%% 
%%%%%%%%%%%%%%%%%%%%%
%%%%%%%%%%%%%%%%%%%%%
%%%%%%%%%%%%%%%%%%%%%


\begin{thebibliography}{10}
	
	\bibitem{warner1996nematic}
	Warner M, Terentjev E.
	\newblock Nematic elastomers—a new state of matter?
	\newblock Progress in Polymer Science. 1996;21(5):853-91.
	
	\bibitem{white2015programmable}
	White TJ, Broer DJ.
	\newblock Programmable and adaptive mechanics with liquid crystal polymer networks and elastomers.
	\newblock Nature materials. 2015;14(11):1087-98.
	
	\bibitem{fuchi2015topology}
	Fuchi K, Ware TH, Buskohl PR, Reich GW, Vaia RA, White TJ, et~al.
	\newblock Topology optimization for the design of folding liquid crystal elastomer actuators.
	\newblock Soft matter. 2015;11(37):7288-95.
	
	\bibitem{babaei2021torque}
	Babaei M, Gao J, Clement A, Dayal K, Shankar MR.
	\newblock Torque-dense photomechanical actuation.
	\newblock Soft Matter. 2021;17(5):1258-66.
	
	\bibitem{ahn2016photoinduced}
	Ahn Sk, Ware TH, Lee KM, Tondiglia VP, White TJ.
	\newblock Photoinduced topographical feature development in blueprinted azobenzene-functionalized liquid crystalline elastomers.
	\newblock Advanced Functional Materials. 2016;26(32):5819-26.
	
	\bibitem{ware2016localized}
	Ware TH, Biggins JS, Shick AF, Warner M, White TJ.
	\newblock Localized soft elasticity in liquid crystal elastomers.
	\newblock Nature communications. 2016;7(1):10781.
	
	\bibitem{ahn2019bioinspired}
	Ahn C, Liang X, Cai S.
	\newblock Bioinspired design of light-powered crawling, squeezing, and jumping untethered soft robot.
	\newblock Advanced materials technologies. 2019;4(7):1900185.
	
	\bibitem{ware2015voxelated}
	Ware TH, McConney ME, Wie JJ, Tondiglia VP, White TJ.
	\newblock Voxelated liquid crystal elastomers.
	\newblock Science. 2015;347(6225):982-4.
	
	\bibitem{babaei2017steering}
	Babaei M, Clement JA, Dayal K, Shankar MR.
	\newblock Steering with light: indexable photomotility in liquid crystalline polymers.
	\newblock RSC advances. 2017;7(83):52510-6.
	
	\bibitem{dradrach2023light}
	Dradrach K, Zmy{\'s}lony M, Deng Z, Priimagi A, Biggins J, Wasylczyk P.
	\newblock Light-driven peristaltic pumping by an actuating splay-bend strip.
	\newblock Nature Communications. 2023;14(1):1877.
	
	\bibitem{javed2022programmable}
	Javed M, Corazao T, Saed MO, Ambulo CP, Li Y, Kessler MR, et~al.
	\newblock Programmable Shape Change in Semicrystalline Liquid Crystal Elastomers.
	\newblock ACS Applied Materials \& Interfaces. 2022;14(30):35087-96.
	
	\bibitem{lewis2023programming}
	Lewis KL, Herbert KM, Matavulj VM, Hoang JD, Ellison ET, Bauman GE, et~al.
	\newblock Programming Orientation in Liquid Crystalline Elastomers Prepared with Intra-Mesogenic Supramolecular Bonds.
	\newblock ACS Applied Materials \& Interfaces. 2023.
	
	\bibitem{CLEMENT2021101362}
	Clement A, Babaei M, Phadikar J, Lee DW, Skandani AA, Shankar MR.
	\newblock Complexity from simplicity: Confinement directs morphogenesis and motility in nematic polymers.
	\newblock Extreme Mechanics Letters. 2021;47:101362.
	\newblock Available from: \url{https://www.sciencedirect.com/science/article/pii/S2352431621001127}.
	
	\bibitem{giudici2022curvature}
	Giudici A, Biggins JS.
	\newblock Curvature-driven instabilities in thin active shells.
	\newblock Royal Society Open Science. 2022;9(10):220487.
	
	\bibitem{kadapa2021advantages}
	Kadapa C, Li Z, Hossain M, Wang J.
	\newblock On the advantages of mixed formulation and higher-order elements for computational morphoelasticity.
	\newblock Journal of the Mechanics and Physics of Solids. 2021;148:104289.
	
	\bibitem{agostiniani2017dimension}
	Agostiniani V, DeSimone A.
	\newblock Dimension reduction via {$\Gamma$}-convergence for soft active materials.
	\newblock Meccanica. 2017;52(14):3457-70.
	
	\bibitem{Singh2022}
	Singh H, Virga E.
	\newblock A Ribbon Model for Nematic Polymer Networks.
	\newblock Journal of Elasticity. 2022 06:1-22.
	
	\bibitem{D0SM00642D}
	Ozenda O, Sonnet AM, Virga EG.
	\newblock A blend of stretching and bending in nematic polymer networks.
	\newblock Soft Matter. 2020;16:8877-92.
	\newblock Available from: \url{http://dx.doi.org/10.1039/D0SM00642D}.
	
	\bibitem{bartels2022modeling}
	Bartels S, Griehl M, Keck J, Neukamm S.
	\newblock Modeling and simulation of nematic LCE rods.
	\newblock arXiv preprint arXiv:220515174. 2022.
	
	\bibitem{agostiniani2011gamma}
	Agostiniani V, DeSimone A.
	\newblock {$\Gamma$}-convergence of energies for nematic elastomers in the small strain limit.
	\newblock Continuum Mechanics and Thermodynamics. 2011;23(3):257-74.
	
	\bibitem{agostiniani2017shape}
	Agostiniani V, DeSimone A, Koumatos K.
	\newblock Shape programming for narrow ribbons of nematic elastomers.
	\newblock Journal of Elasticity. 2017;127(1):1-24.
	
	\bibitem{agostiniani2020rigorous}
	Agostiniani V, DeSimone A.
	\newblock Rigorous derivation of active plate models for thin sheets of nematic elastomers.
	\newblock Mathematics and Mechanics of Solids. 2020;25(10):1804-30.
	
	\bibitem{zajac1962stability}
	Zajac E.
	\newblock Stability of two planar loop elasticas.
	\newblock Journal of Applied Mechanics. 1962;29(1):136-42.
	
	\bibitem{green1936equilibrium}
	Green AE.
	\newblock The equilibrium and elastic stability of a thin twisted strip.
	\newblock Proceedings of the Royal Society of London Series A-Mathematical and Physical Sciences. 1936;154(882):430-55.
	
	\bibitem{green1937elastic}
	Green AE.
	\newblock The elastic stability of a thin twisted strip—II.
	\newblock Proceedings of the Royal Society of London Series A-Mathematical and Physical Sciences. 1937;161(905):197-220.
	
	\bibitem{purohit2008plectoneme}
	Purohit PK.
	\newblock Plectoneme formation in twisted fluctuating rods.
	\newblock Journal of the Mechanics and Physics of Solids. 2008;56(5):1715-29.
	
	\bibitem{steigmann2008two}
	Steigmann DJ.
	\newblock Two-dimensional models for the combined bending and stretching of plates and shells based on three-dimensional linear elasticity.
	\newblock International Journal of Engineering Science. 2008;46(7):654-76.
	
	\bibitem{korte2011triangular}
	Korte A, Starostin E, Van Der~Heijden G.
	\newblock Triangular buckling patterns of twisted inextensible strips.
	\newblock Proceedings of the Royal Society A: Mathematical, Physical and Engineering Sciences. 2011;467(2125):285-303.
	
	\bibitem{fosdick2016mechanics}
	Fosdick R, Fried E.
	\newblock The mechanics of ribbons and M{\"o}bius bands.
	\newblock Springer; 2016.
	
	\bibitem{yu2019bifurcations}
	Yu T, Hanna J.
	\newblock Bifurcations of buckled, clamped anisotropic rods and thin bands under lateral end translations.
	\newblock Journal of the Mechanics and Physics of Solids. 2019;122:657-85.
	
	\bibitem{dondl2023efficient}
	Dondl P, Luo Y, Neukamm S, Wolff-Vorbeck S.
	\newblock Efficient uncertainty quantification for mechanical properties of randomly perturbed elastic rods.
	\newblock arXiv preprint arXiv:230408785. 2023.
	
	\bibitem{radisson2023dynamic}
	Radisson B, Kanso E.
	\newblock Dynamic behavior of elastic strips near shape transition.
	\newblock Physical Review E. 2023.
	
	\bibitem{radisson2023elastic}
	Radisson B, Kanso E.
	\newblock Elastic snap-through instabilities are governed by geometric symmetries.
	\newblock arXiv preprint arXiv:230212152. 2023.
	
	\bibitem{antman2005problems}
	Antman SS.
	\newblock Problems in nonlinear elasticity.
	\newblock Springer; 2005.
	
	\bibitem{audoly2000elasticity}
	Audoly B, Pomeau Y.
	\newblock Elasticity and geometry; 2000.
	
	\bibitem{seffen2000folding}
	Seffen K, You Z, Pellegrino S.
	\newblock Folding and deployment of curved tape springs.
	\newblock International Journal of Mechanical Sciences. 2000;42(10):2055-73.
	
	\bibitem{kumar2023asymptotic}
	Kumar A, Audoly B, Lestringant C.
	\newblock Asymptotic derivation of a higher-order one-dimensional model for tape springs.
	\newblock Philosophical Transactions of the Royal Society A. 2023;381(2244):20220028.
	
	\bibitem{audoly2016buckling}
	Audoly B, Seffen KA.
	\newblock Buckling of naturally curved elastic strips: the ribbon model makes a difference.
	\newblock Journal of Elasticity. 2015;119:293-320.
	
	\bibitem{moulton2020morphoelastic}
	Moulton DE, Lessinnes T, Goriely A.
	\newblock Morphoelastic rods III: Differential growth and curvature generation in elastic filaments.
	\newblock Journal of the Mechanics and Physics of Solids. 2020;142:104022.
	
	\bibitem{GUINOT201273}
	Guinot F, Bourgeois S, Cochelin B, Blanchard L.
	\newblock A planar rod model with flexible thin-walled cross-sections. Application to the folding of tape springs.
	\newblock International Journal of Solids and Structures. 2012;49(1):73-86.
	\newblock Available from: \url{https://www.sciencedirect.com/science/article/pii/S002076831100312X}.
	
	\bibitem{anderson1999continuum}
	Anderson DR, Carlson DE, Fried E.
	\newblock A continuum-mechanical theory for nematic elastomers.
	\newblock Journal of Elasticity. 1999;56:33-58.
	
	\bibitem{desimone2009elastic}
	DeSimone A, Teresi L.
	\newblock Elastic energies for nematic elastomers.
	\newblock The European Physical Journal E. 2009;29:191-204.
	
	\bibitem{le2023numerical}
	Le~Cl{\'e}zio H, Lestringant C, Kochmann DM.
	\newblock A numerical two-scale approach for nonlinear hyperelastic beams and beam networks.
	\newblock International Journal of Solids and Structures. 2023;276:112307.
	
	\bibitem{BERG20145849}
	Berg GJ, McBride MK, Wang C, Bowman CN.
	\newblock New directions in the chemistry of shape memory polymers.
	\newblock Polymer. 2014;55(23):5849-72.
	\newblock Shape Memory and Shape Morphing Polymers.
	\newblock Available from: \url{https://www.sciencedirect.com/science/article/pii/S0032386114006612}.
	
	\bibitem{vlassov1962pieces}
	Vlassov B.
	\newblock Pi{\`e}ces longues en voiles minces, traduit du Russe par G.
	\newblock Smirnoff, Eyrolles. 1962.
	
	\bibitem{seedPod}
	Armon S, Efrati E, Kupferman R, Sharon E.
	\newblock Geometry and Mechanics in the Opening of Chiral Seed Pods.
	\newblock Science. 2011;333(6050):1726-30.
	\newblock Available from: \url{https://www.science.org/doi/abs/10.1126/science.1203874}.
	
	\bibitem{coleman2000theory}
	Coleman BD, Swigon D.
	\newblock Theory of supercoiled elastic rings with self-contact and its application to DNA plasmids.
	\newblock Journal of elasticity and the physical science of solids. 2000;60:173-221.
	
	\bibitem{van2003instability}
	Van~der Heijden G, Neukirch S, Goss V, Thompson J.
	\newblock Instability and self-contact phenomena in the writhing of clamped rods.
	\newblock International Journal of Mechanical Sciences. 2003;45(1):161-96.
	
	\bibitem{friesecke2006hierarchy}
	Friesecke G, James RD, M{\"u}ller S.
	\newblock A hierarchy of plate models derived from nonlinear elasticity by gamma-convergence.
	\newblock Archive for rational mechanics and analysis. 2006;180:183-236.
	
	\bibitem{coifman2006diffusion}
	Coifman RR, Lafon S.
	\newblock Diffusion maps.
	\newblock Applied and computational harmonic analysis. 2006;21(1):5-30.
	
\end{thebibliography}
\end{document}